\documentclass{aa}
\usepackage{url} 
\usepackage{graphicx}
\usepackage{txfonts}
\usepackage{multirow}
\usepackage{amsmath}
\usepackage{enumitem}
\usepackage{relsize}
\usepackage[colorlinks=true,allcolors=blue]{hyperref}

\usepackage{xcolor}
\newcommand{\ma}[1]{\text{#1}}
\newcommand{\s}{{\mathlarger{\chi}}}
\newcommand{\Lsigma}{{\mathlarger{\sigma}}}

\newcommand{\sQ}{\ensuremath{\s_{\ma{G}}^\ma{4Q}}}

\newcommand{\correction}[1]{\textcolor{black}{#1}}
\begin{document} 

\title{The Bi-O-edge wavefront sensor}
\subtitle{How Foucault-knife-edge variants can boost eXtreme Adaptive Optics}
\author{C. V\'erinaud
          \inst{1}
          \and
          C.T. H\'eritier \inst{1,2,3}
          \and
          M. Kasper\inst{1}
          \and
          M. Tallon\inst{4}
          }
\institute{European Southern Observatory (ESO),
              Karl Schwarzschild-str 2, D-85748 Garching-bei-Muenchen\- Germany, \
              \email{cverinau@eso.org}
              \and
         DOTA, ONERA, F-13661 Salon cedex Air - France
         \and
         Aix Marseille Univ, CNRS, CNES, LAM, Marseille, France
         \and
             Centre de Recherche Astrophysique de Lyon (CRAL), 
             9 avenue Charles Andr\'e F-69230
             Saint Genis-Laval - France}
\date{Received 14 Apr 2023; accepted  09  Aug 2023, }
 
\abstract
  {Direct detection of {exoplanets} around nearby stars requires advanced Adaptive Optics (AO) systems. High order systems are needed to reach high Strehl Ratio (SR) in near infrared and optical wavelengths on future Giant Segmented Mirror Telescopes (GSMTs). Direct detection of faint exoplanets with the ESO's ELT will require some tens of thousand of correction modes. Resolution and sensitivity of the wavefront sensor (WFS) are key requirements for this science case. We present a new class of WFSs, the 'Bi-Orthogonal Foucault-knife-edge Sensors' (or Bi-O-edge), that is directly inspired by the Foucault knife edge test \citep{Foucault_1859AnPar_5_197F}. The  idea consists of using a beam-splitter producing two foci, each of which is sensed by an edge with orthogonal direction to the other.}
  {We describe two implementation concepts: The Bi-O-edge sensor can be realised with a sharp edge and a tip-tilt modulation device (sharp Bi-O-edge) or with a smooth gradual transmission over a 'grey' edge (grey Bi-O-edge).
  A comparison between the  Bi-O-edge concepts  and the 4-sided classical Pyramid Wavefront Sensor (PWS)  gives some  important insights into the nature of the measurements.}
  {We compute analytically the photon noise error propagation and we compare the results to end-to-end simulations of a closed loop AO system.}
  { Our analysis shows that the sensitivity gain of the Bi-O edge with respect to the PWS depends on the system configuration. The gain is a function of the number of control modes and the modulation angle. We found that for the sharp Bi-O-edge, the gain in reduction of propagated photon noise variance approaches a  theoretical factor of $2$ for a large number of control modes and small modulation angle, meaning that the sharp Bi-O-edge only needs half of the photons of the PWS to reach similar measurement accuracy. On the contrary, the PWS is twice more sensitive than the Bi-O-edge, in the case of very low order correction and/or large modulation angles. Preliminary end-to-end simulations illustrate some of the results. The grey version of the Bi-O-edge opens the door to advanced amplitude filtering replacing the need for a tip-tilt modulator while keeping the same dynamic range. We show that an additional factor of $2$ in reduction of propagated photon noise variance can be obtained for high orders such that the theoretical maximum gain of a factor of $4$ in photon efficiency can be obtained. A  diffractive Fourier model that includes accurately the effect of modulation and control modes, shows that for the ELT XAO/PCS system configuration, the overall gain will well exceed one magnitude in guide star brightness when compared to the modulated PWS.}
  {We conclude that the Bi-O edge is an excellent  candidate sensor for future very high order Adaptive Optics systems, in particular on GSMTs.}


\maketitle

\section{Introduction}
 High-contrast direct imaging (HCI) of exoplanets from the ground is one the most demanding applications of Adaptive Optics (AO). Current HCI instruments such as SPHERE \citep{SPHERE_2019A&A_631A.155B}, GPI \citep{GPI_2018SPIE10703E_0KM}, SCExAO \citep{SCEXAO_2015PASP_127_890J}, KPIC \citep{KPIC_2018SPIE10703E_06M} or MagAO-X \citep{MAG-AOX_2018SPIE10703E_09M} installed on 8-m class telescopes reach high contrast sensitivities which led to the discovery of several young giant planets \citep[e.g.][]{Macintosh_Planet_2015Sci_350_64M, Keppler_Planet_2018A&A_617A_44K, Lagrange_Planet_2010Sci_329_57L}.
 The direct imaging method has also allowed powerful characterisation of the planetary atmospheres through direct spectroscopy, returning not only effective temperatures and surface gravities, but also detections of molecular species which provide basic estimates of the atmospheric compositions \citep[e.g.][]{Konopacky_spectro_2013Sci_339.1398K}. 

A major achievement of exoplanetary science in the last several years is the determination that low-mass planets are common \citep{Dressing_2013ApJ_767_95D}, and the identification of numerous more such objects is expected to proceed in the coming years. Looking forward into the 2030's and beyond, the new generation of giant 30- to 40-m class telescopes (ELT, GMT, TMT) should be capable of detecting and characterising such small planets of Earth to sub-Neptune size around the closest M-dwarfs and even when located in the habitable zone \citep{Kasper_2021Msngr.182_38K}.

HCI instruments typically combine eXtreme AO \citep[XAO,][]{Guyon_XAO_2005ApJ_629_592G}, coronagraphy \citep{Mawet_corono_2012SPIE.8442E_04M} and quasi-static speckle control \citep[e.g.][]{Give_on_2007SPIE.6691E_0AG} as well as advanced post-processing \citep[e.g.][] {Marois_2006ApJ_641_556M,Hoeijmakers2018A&A_617A.144H}. Such concepts promise to reduce speckle noise effectively to the level where instruments are limited by photon noise of the XAO residual halo of the coronagraphic Point Spread Function (PSF).

 In HCI limited by photon noise, the observing time is proportional to the square of the Signal-to-Noise ratio (S/N). The latter being proportional to the Strehl Ratio (SR), it becomes critical for HCI to maximize the SR and minimize the residual halo over the control radius of the XAO deformable mirror (DM). 

For instance, the SPHERE instrument was designed to detect exoplanets with atmospheres containing methane  at the 1.65 $\mu m$ absorption feature in  H band. To do so, the primary top-level requirement of SPHERE eXtreme Adaptive Optics (XAO) SAXO \citep{Fusco_SAXO_2014SPIE.9148E_1UF} is to reach an almost perfect light concentration in the core of the PSF at the observing wavelength with a SR of $90~\%$ in H band. This was achieved by setting the resolution of the sensor to $20~cm$ sampling of the telescope aperture. The ultimate goal of the ELT Planetary Camera Spectrograph (PCS, \citeauthor{Kasper_2021Msngr.182_38K},\citeyear{Kasper_2021Msngr.182_38K}) is to detect bio-markers in exoplanets atmosphere, e.g., the A-band of molecular oxygen at around 760 $nm$. Reaching a high SR becomes more difficult at shorter wavelengths and will ultimately be limited by the DM fitting error. Hence the PCS XAO system must push AO to its limits and calls for the most sensitive WFS in order to minimize the residual halo.

In this paper, we revisit the concept of the 'two-sided PWS' proposed by \cite{Phillion_2006SPIE.6272E_28P}. We generalise the concept and propose new optical implementations. To underline the  nature of the focal plane elements, we name the concept the 'Bi-Orthogonal-Foucault-knife-edge' sensor (short name: Bi-O-edge). We compare its properties to a reference, the well known PWS  \citep{Ragazzoni_Pyramid_1996JMOp_43_289R}. This sensor is now well established with many systems producing science on sky, e.g. at the Large Binocular Telescope \citep{Esposito_Planet_2013A&A_549A_52E}, on SUBARU \citep{SCEXAO_2015PASP_127_890J} , the Keck telescope \citep{Bond_Pyramid_Keck_2020JATIS_6c9003B}, the Magellan Clay Telescope \citep{MAG-AOX_2018SPIE10703E_09M}  and  projects in development for the ELT \citep[e.g.][] {MICADO_2018SPIE10703E_13C,METIS_2018SPIE10703E_14B,HARMONI_2020arXiv200307228S}, and the TMT \citep{TMT_PYRAMID_LGS_2018SPIE10703E_3VC}.
 
 After the presentation of the Wave-Front sensing context in Sect.\ref{sec:WFS context}, we analyse the Fourier filtering properties of a Foucault Knife Edge (FKE) in Sect. \ref{sect_FKE}.  The two flavors of Bi-O-edge (sharp and grey) are presented in Sect. \ref{sec:BI-O-EDGE_concept}. In Section \ref{sec:sensitivity_noise_prop}, the FKE properties are used to derive the PWS and Bi-O-edge sensitivities and noise propagation. In Section \ref{sec:Modal_Analysis}, we use a modal approach to compare more accurately the performance for both Bi-O-edge and PWS and show the dependence with the number of corrected modes as well as closed loop simulations obtained with end-to-end models. A Fourier model for Fourier Filtering WFS (FF-WFS) is given in appendix \ref{sec:C_model}. This model is called the Convolutional model (C-model, hereafter) and is used for the analytical developments presented in this paper.

\section{Wave-front sensing context}
\label{sec:WFS context}

The WFS is an essential part of an astronomical AO system.
The increasing needs for high precision, high sensitivity and very large number of degrees of freedom (DoF) calls for a careful study of the WFS properties. During early days of AO, the Lateral Shearing Interferometer (LSI) was the most commonly used sensor \citep{ROUSSET_1999aoa_book_91R}. It is interesting to notice that this slope sensor required two channels like the Bi-O-edge does, one for each orthogonal wave-front derivative component. 

Since ADONIS, the first workhorse astronomical AO instrument \citep{BEUZIT_ADONIS_1997ExA_7_285B}, the Shack-Hartmann Sensor (SHS) became the most used WFS in AO. The success of the SHS was largely based on its conceptual simplicity, achromaticity, and large linear range \citep{ROUSSET_1999aoa_book_91R}. In contrast to the LSI, the SHS was maximising the flux sensitivity and simplifying the opto-mechanical concept. 

The PWS \citep{Ragazzoni_Pyramid_1996JMOp_43_289R}, represented a giant leap in sensitivity at the expense of a slightly larger complexity and smaller dynamic range compared to the SHS. To cope with the issues of dynamic range, the PWS sensor is generally coupled to a tip-tilt modulated mirror that allows to improve the linear range at the cost of some sensitivity. 
The sensitivity gain of the PWS over the SHS can be tremendous for high order systems and was studied extensively \citep[see][]{RAGAZZONI_FARINATO_1999A&A_350L_23R,ESPOSITO_PYR_2001A&A_369L_9E,Verinaud_2005MNRAS.357L_26V,Guyon_XAO_2005ApJ_629_592G}

The class of FF-WFS is a generalization of the \correction{PWS} concept first introduced and studied by \cite{FAUVARQUE_OPTICS_2016Optic_3.1440F}. Using a few hypotheses, a theoretical formalism based on a C-model has been developed and allows one to derive analytical transfer functions (TFs) depending on the filtering mask property. 

\begin{figure}[h!]
\centering\includegraphics[width=\linewidth]{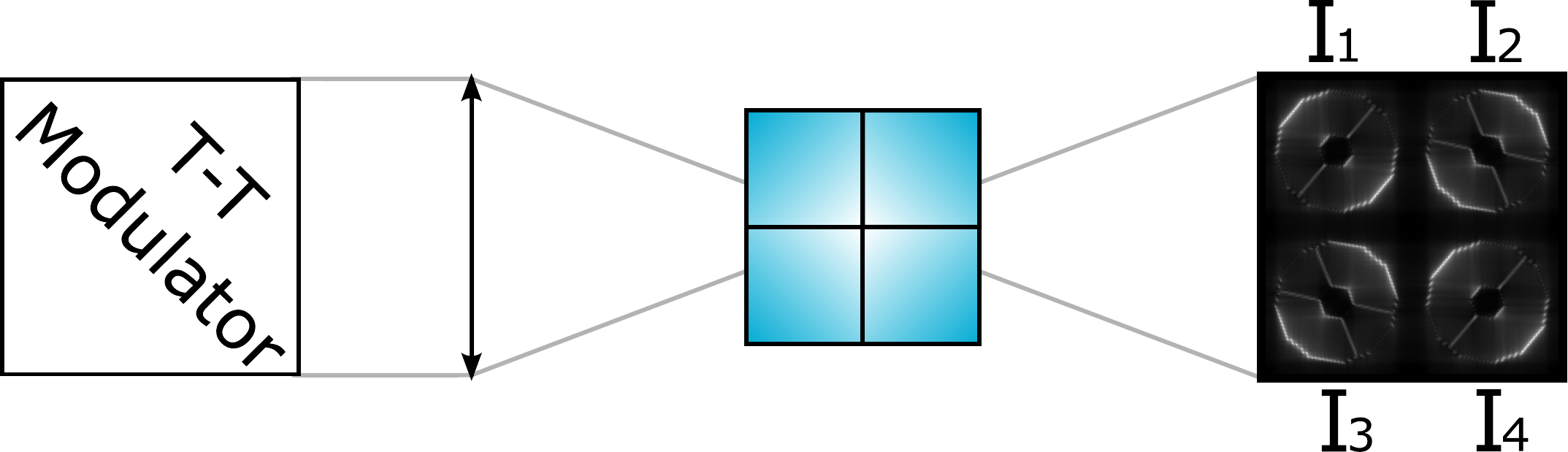}
\caption{Schematic view of the concept of modulated PWS sensor with a refractive Pyramidal prism. \correction{A Tip-Tilt modulation mirror moves the focal spot over a 4-facet Pyramid. The  signal is obtained by integrating  the intensity on a pupil plane detector during a modulation cycle.}}
\label{fig:PYR_MOD}
\end{figure}

Throughout this study, the PWS, as the most common FF-WFS (see Fig. \ref{fig:PYR_MOD}), is used as a reference for the exploration of the Bi-O-edge properties. Only \correction{circularly} modulated PWSs are considered. While the PWS can be operated without modulation \citep[e.g.][]{Costa_modulation_2005ApOpt_44_60C,Jalo_Pyramid_2022arXiv220507554N}, the very small linear range of the non-modulated PWS makes it hard to operate in practice.

Interestingly, slope sensor concepts based on two channels  with focal amplitude masks have been proposed \correction{\citep[e.g.][]{Horwitz_1994SPIE.2201_496H, Oti_OD_2005MNRAS.360.1448O, Haffert_GODS_2016OExpr_2418986H, HENAULT_2020ERExp...2a5042H}} and share some practical implementation solutions with one of the variants of the Bi-O-edge presented in this paper. 

In general, WFSs can be categorised into two families: the geometric and diffractive WFSs. Geometric WFSs are characterised by a large dynamic range and a low sensitivity (the SHS) while the diffractive WFS like the PWS offer a high sensitivity but are usually associated to a smaller dynamic range. The optimum choice depends on the scientific objective of the AO system. In this paper, we propose to study the Bi-O-edge, a diffractive WFS concept offering unprecedented sensitivity for the high-contrast XAO.
\section{The Foucault Knife Edge as a Fourier Filtering Sensor}
\label{sect_FKE}
The FKE Test \citep{Foucault_1859AnPar_5_197F} is commonly used in Astronomy to quantify the radius of curvature of optical devices by masking part of the ray-light in a pre and post-focal plane. From a wave-front sensing perspective, if the mask is located in the focal plane, it can be seen as the most elementary FF-WFS. In \cite{Verinaud_Pyramid_2004OptCo.233_27V}, a mono-dimensional model of a tip-tilt modulated FKE was used as a simplified model of the PWS. In this section, we generalise this model to $2$ dimensions in order to highlight some remarkable properties.

\subsection{Nature of the measurements of a Foucault Knife Edge}
\label{sec:NATURE_FKE}
\begin{figure}[h]
    \centering
    \begin{tabular}{ccc}
        \textsf{\footnotesize{Wave-Front}}& \textsf{\footnotesize{PSF}}&\textsf{\footnotesize{Pupil Intensity}}  \\
        \includegraphics[width = 0.14\textwidth]{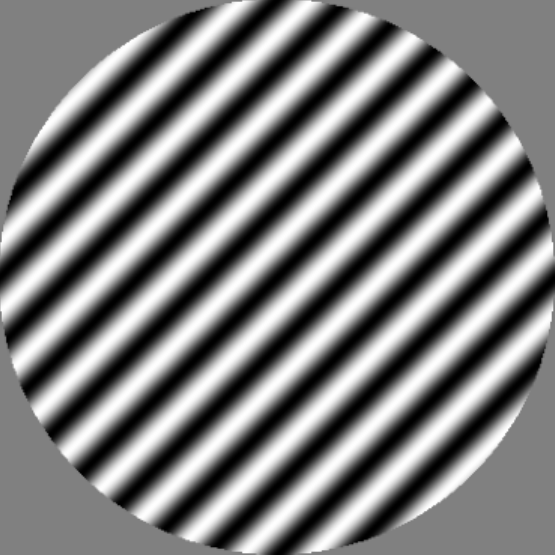}&
        \includegraphics[width = 0.14\textwidth]{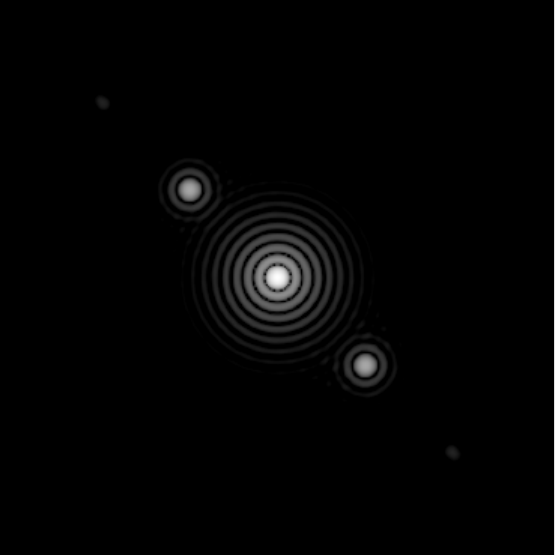}&
        \includegraphics[width = 0.14\textwidth]{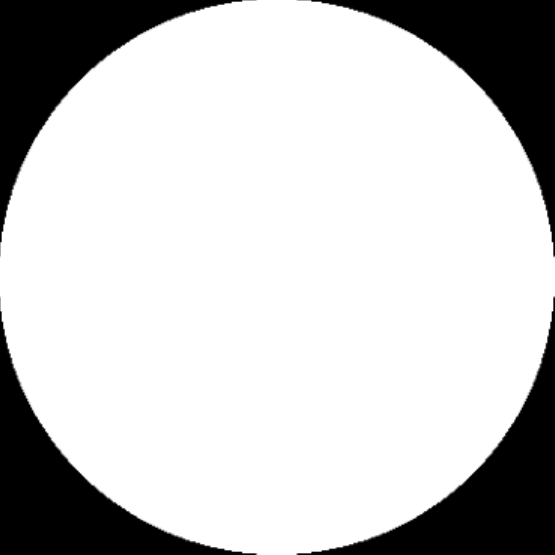}
    \end{tabular}
    \caption{Focal (Center) and Pupil plane (Right) images corresponding to a pure spatial frequency wave-front (Left).}
    \label{fig:FKE_no_mask}
\end{figure}

Let's consider a pure sinusoidal aberration of standard deviation $\sigma_\phi$ as a test wave-front with spatial frequency $w_0$ and let $r$ be the variable along the sinusoidal function axis. We define the measurement as the meta-intensity $mI$, which is a linear combination of pupil intensity maps (with null WF reference maps subtracted).  The sensitivity $\s^{mI}$ in the small phase regime is given by \citep{FAUVARQUE_OPTICS_2016Optic_3.1440F}:
\begin{equation}
\label{eq:SENSITIVITY_DEF}
\s^{mI} = \frac{|| mI(\phi)||_2}{\sigma_{\phi}}
\end{equation}
where $||\cdot||_2$ is the L2 norm.
 In the small phase regime, this WF creates two symmetric speckles in the focal plane at a distance from the core that depends on the spatial frequency \citep{Malbet_Dark_Hole_1995PASP_107_386M}. The spatial frequency is chosen to be high enough to separate the speckles from the PSF core, as illustrated in Fig. \ref{fig:FKE_no_mask}. 
In the corresponding pupil plane, a uniform pupil is visible with intensity $I_0$ that corresponds to the square modulus of the electromagnetic field. In this situation, both speckles are actually interfering with the core of the PSF but the resulting interference fringes are masking each other's impact in the pupil plane. 

\begin{figure}[h]
    \centering
    \begin{tabular}{ccc}
        \textsf{\footnotesize{Wave-Front}}& \textsf{\footnotesize{PSF}}&\textsf{\footnotesize{Pupil Intensity}}  \\
        \includegraphics[width = 0.14\textwidth]{FIGS/fourier_modes_WF.pdf}&
        \includegraphics[width = 0.14\textwidth]{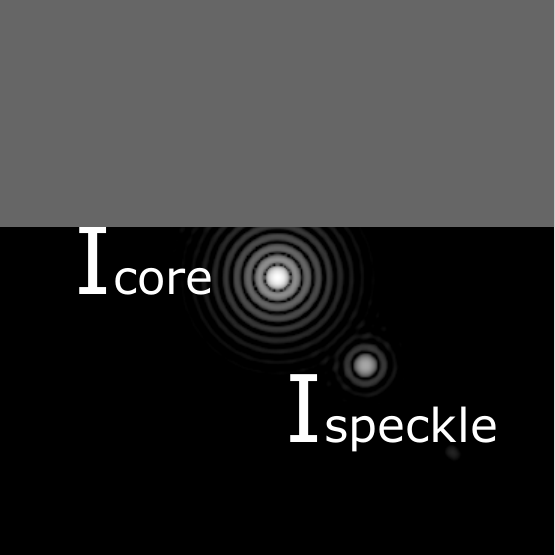}&
        \includegraphics[width = 0.14\textwidth]{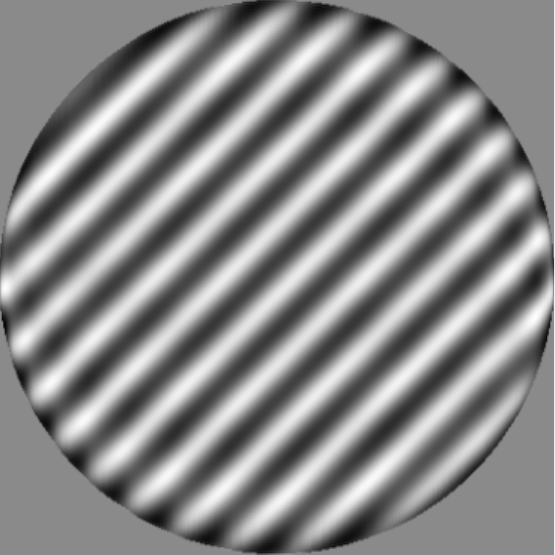}
    \end{tabular}

    \caption{Focal (Center) and Pupil plane (Right) images corresponding to a pure spatial frequency wave-front (Left) in presence of an amplitude mask.}
    \label{fig:FKE_maskl}
\end{figure}

If an amplitude mask is now added in the focal plane to mask one of the two speckles, as illustrated in Figure \ref{fig:FKE_maskl}, it operates a filtering of one of the satellite speckle. This operation allows to reveal the interference pattern between the non masked speckle with intensity $I_{speckle}$ and the core of the PSF of intensity $I_{core}$. 

The resulting fringes amplitude $I^{CS}_\sigma$ is:
\begin{equation}
I^{CS}_\sigma = 2\sqrt{I_{core}I_{speckle}}
\label{eq:I_CORE_SPECKLE}
\end{equation}

The magnitude and phase of the fringes corresponds to the Hilbert transform \citep{Correia_Model_Pyramid_2020MNRAS.495.4380C} of the incoming wave-front ($\pm \pi/2$ dephasing depending on which speckle interferes). 
In the small phase regime, one can approximate the intensity in the core and the intensity in one speckle with: 
\begin{equation}
\label{Ics_approx}
I_{core}\approx (1-\sigma_{\phi}^2) I_0 \\
I_{speckle}\approx \frac{\sigma_{\phi}^2}{2}I_0
\end{equation}
We can express the meta-intensity \correction{corresponding to the Core-Speckle (CS) interference} as:
\begin{equation}
\label{mI_f_cs_approx}
 mI^{CS}(\phi) \approx I^{CS}_\sigma \sin(2 \pi w_0 r+\theta)/I_0
\end{equation}
Noting that $||\sin(2 \pi w_0 r+\theta)||_2=\frac{1}{\sqrt{2}}$, and keeping only the first order in $\sigma_{\phi}$, we find that $|| mI^{CS}(\phi)||_2=\sigma_{\phi}$.

Hence, the sensitivity corresponding to the \correction{CS} meta-intensity is:
\begin{equation}
\s^{CS} = \frac{|| mI^{CS}(\phi)||_2} {\sigma_{\phi}} = 1
    \label{eq:STD_I_fringe}
\end{equation}
In Section \ref{sec:sensitivity_noise_prop}, the sensitivities of the WFS concepts considered will be expressed as the CS sensitivity $\s^{CS}$ multiplied by an efficiency factor depending on both the Fourier filter and the modulation path.

This concept of filtering is central to all the FF-WFS variants and can easily be generalized even in presence of tip-tilt modulation, as long as the satellite speckles are properly masked for at least some part of the modulation path. Moreover, since the fringe pattern depends only on the relative positions of the core and the speckle, modulation does not blur the fringes.

The tip-tilt modulation was historically designed to reproduce the measurement of a quad-cell sensor to provide WFS measurements that can be associated to the gradient of the input WF \citep{Ragazzoni_Pyramid_1996JMOp_43_289R}. At the cost of sensitivity, this operation allows to increase significantly the dynamic range on the low order modes (for spatial frequencies under the modulation radius).
This result has been confirmed in \cite{Verinaud_Pyramid_2004OptCo.233_27V} using a simplified model of the PWS (\textit{e.g.} a single FKE). It demonstrated that depending on the spatial frequency and tip-tilt modulation radius, the nature of the measurement can be associated either to the gradient of the wave-front or to its Hilbert transform. 

\subsection{Orthogonal {Foucault Knife Edge}s with modulation}
\label{sec:DOUBLE_FKE}

One straightforward generalisation of the FKE mono-dimensional model presented in \cite{Verinaud_Pyramid_2004OptCo.233_27V} is to consider the information on the WF provided by two distinct orthogonal FKEs with a linear and uniform tip-tilt modulation orthogonal to each edge . Under this assumption and following the results presented in \cite{Verinaud_Pyramid_2004OptCo.233_27V}, we can rank the modes depending on their Fourier components $(u,v)$ as follows:
\begin{itemize}
    \item G-modes (measured like Gradient): $\sqrt{u^2+v^2}<r_{mod}/D$
    \item H-modes (measured like Hilbert transforms): $\sqrt{u^2+v^2} > r_{mod}/D $ 
\end{itemize}
where $u$ and $v$ are the spatial frequency coordinates corresponding respectively to $X$ and $Y$. $r_{mod}$ is the radius of the modulation circle expressed in units of $\lambda /D$ where $\lambda$ is the wavelength and $D$ the pupil diameter. For the sake of simplicity, we discard the H-modes with either $|u|<r_{mod}/D$ or $|v|<r_{mod}/D$  since their behaviour is slightly different but do not contribute significantly to the error budget. 

We consider the definition of the slope-like measurements  $\ma{S}_x$ corresponding to two reciprocal Fourier masks of an FKE as defined in  \cite{Verinaud_Pyramid_2004OptCo.233_27V}. We add $\ma{S}_y$ corresponding to an orthogonal FKE. 
For G-modes, the measurements can be \correction{written}:
    \begin{equation}
    \begin{split}
    \ma{S}_x[\phi_G(u,v)] = \frac{iD}{r_{mod}}\cdot u\cdot{\phi_G}(u,v) \\
    \ma{S}_y[\phi_G(u,v)] = \frac{iD}{r_{mod}}\cdot v \cdot {\phi_G}(u,v)
    \end{split}     
    \label{eq:DERIV_FKE}
    \end{equation}
We can trivially note that for any Fourier component with $u \neq v$:
\begin{equation}
        \ma{S}_x[\phi_G(u,v)] \neq \ma{S}_y[\phi_G(u,v)]
        \label{G_mode_property}
\end{equation}
which means that the information in each component is different.
However, for the H-modes $\phi_H(u,v)$, the situation is different as the measurements can be written: 
    \begin{equation}
    \begin{split}
    \ma{S}_x[\phi_H(u,v)] = i\cdot\mathrm{sign}(u)\cdot {\phi_H}(u,v)\\
    \ma{S}_y[\phi_H(u,v)] = i \cdot\mathrm{sign}(v)\cdot {\phi_H}(u,v)
    \end{split}
    \end{equation}
and we have:
\begin{equation}
 \ma{S}_x[\phi_H(u,v)]= \pm \ma{S}_y[\phi_H(u,v)]
        \label{H_mode_property}
\end{equation}
where, this time, each component contains the same information since the difference between $\ma{S}_x$ and $\ma{S}_y$ is only a sign (and this sign depends solely on the signal definition): 

The property of Eq. \ref{H_mode_property} plays a central in our proposition of a new type of WFS that maximizes the sensitivity on the H-modes. In the case of an XAO system with small modulation, H-modes are much more frequent than G-modes  and dominate the overall wave-front error budget.
\subsection{Foucault Knife Edge and Pyramid}
\label{sec:FKE_PYRAMID}
\begin{figure}[h]
\centering
\begin{tabular}{c c}
    Pyramid  & Foucault Knife Edge \\
    \includegraphics[width=0.18\textwidth]{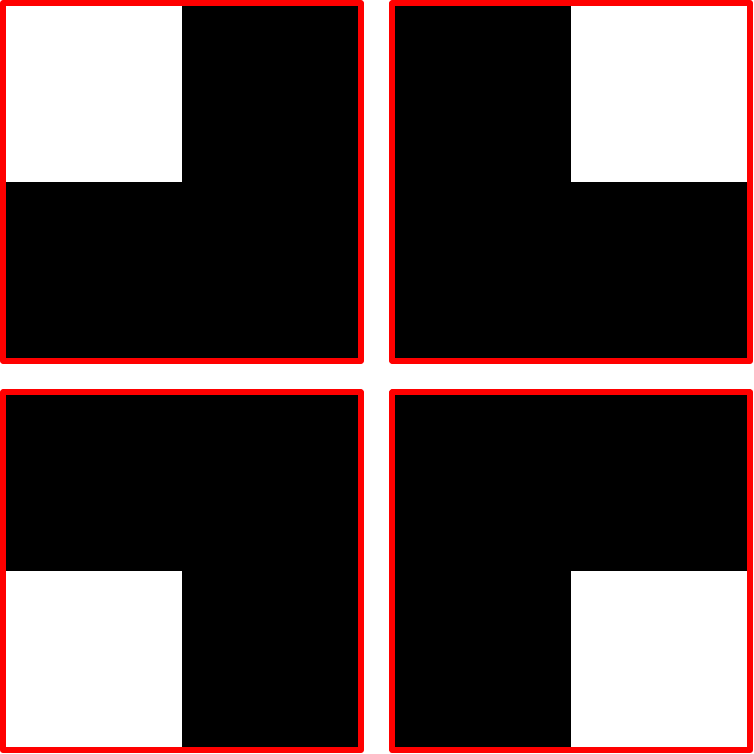} & \includegraphics[width=0.18\textwidth]{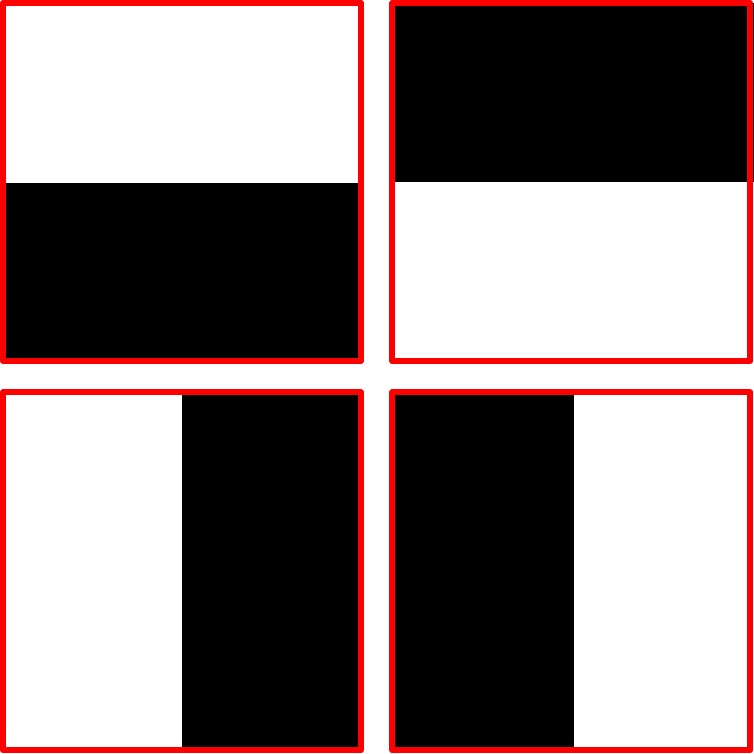}        \end{tabular}    
    \caption{Amplitude masks for each quadrant of the PWS (Left) and equivalent masks for the double FKE (Right).}
    \label{fig:mask_comparison_pyr_fke}
\end{figure}

\begin{figure}[h]
    \centering
    \begin{tabular}{c c}
    PWS&Foucault Knife Edge\\
    \includegraphics[width=0.22\textwidth]{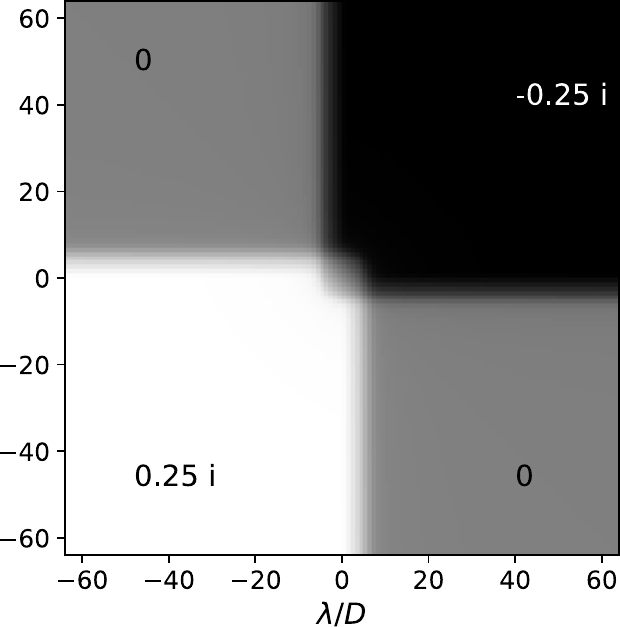}&
    \includegraphics[width=0.22\textwidth]{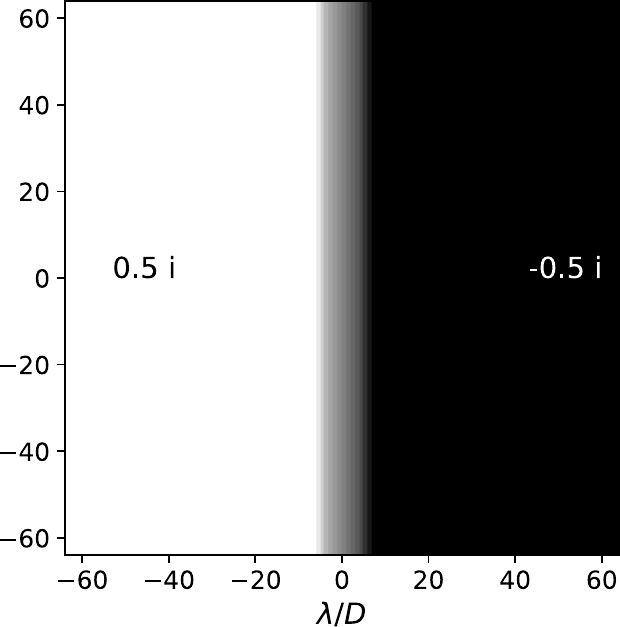}
    \end{tabular}
    \caption{2D Transfer Functions for a single quadrant of the PWS (Left) and for a single FKE (Right). A circular tip-tilt modulation of 3 $\lambda/D$ is considered for both cases. The  TF general definition is given by Eq. \ref{eq:M_TF} and its expression for a single mask is given by Eq. \ref{eq:DEF_TF_MASK}.}
    \label{fig:tf_comparison_pyr_fke}
\end{figure}

\cite{FAUVARQUE_OPTICS_2016Optic_3.1440F} introduced a 2D model for the FF-WFS which uses the filtering masks for each quadrant as provided in Fig. \ref{fig:mask_comparison_pyr_fke}.  
The corresponding 2D TFs between WF and meta-intensity for a single quadrant of the modulated PWS and for a single modulated FKE are provided in Figure \ref{fig:tf_comparison_pyr_fke}.  This figure shows that the TF of a single quadrant of the PWS is characterized by a significant area with null value (top-left and bottom-right) which indicates a blind zone in the Fourier space. As a comparison, the blind zone of the FKE quadrant is much smaller and is concentrated on the frequencies with $u=0$. 
This property as well as the maximum values of the plateaus of the TFs ( $1/4$ for the PWS and $1/2$ for the FKE) that is associated to the sensitivity of the sensors, are explained in Sect. \ref{sec:sensitivity_noise_prop}.

\section{The Bi-Orthogonal Foucault-Knife Edge Sensor concept}
\label{sec:BI-O-EDGE_concept}

\cite{Phillion_2006SPIE.6272E_28P} studied a two-channel non modulated sensor with two orthogonal 'two-sided Pyramids' also called sometimes 'double roof sensor'. This WFS was shown to be a very sensitive 'direct phase sensor' but with a very small dynamic range. For this reason, it was proposed for a second-stage in XAO concept studies like the Planet Formation Imager (PFI on TMT, \cite{Macintosh_PFI_2007amos.confE_62M} and for the Exo-Planet Imaging Camera and Spectrograph (EPICS at ELT, \citeauthor{VERINAUD_EPICS_SYTEM_2010SPIE.7736E_1NV}, \citeyear{VERINAUD_EPICS_SYTEM_2010SPIE.7736E_1NV} ). A full double stage end-to-end simulation of EPICS AO system can be found in \cite{Korkiakoski_BIO_2010SPIE.7736E_43K}.

\subsection{The Sharp Bi-O-Edge}
\label{section_sharp_bio}
\begin{figure}[h]
\centering\includegraphics[width=\linewidth]{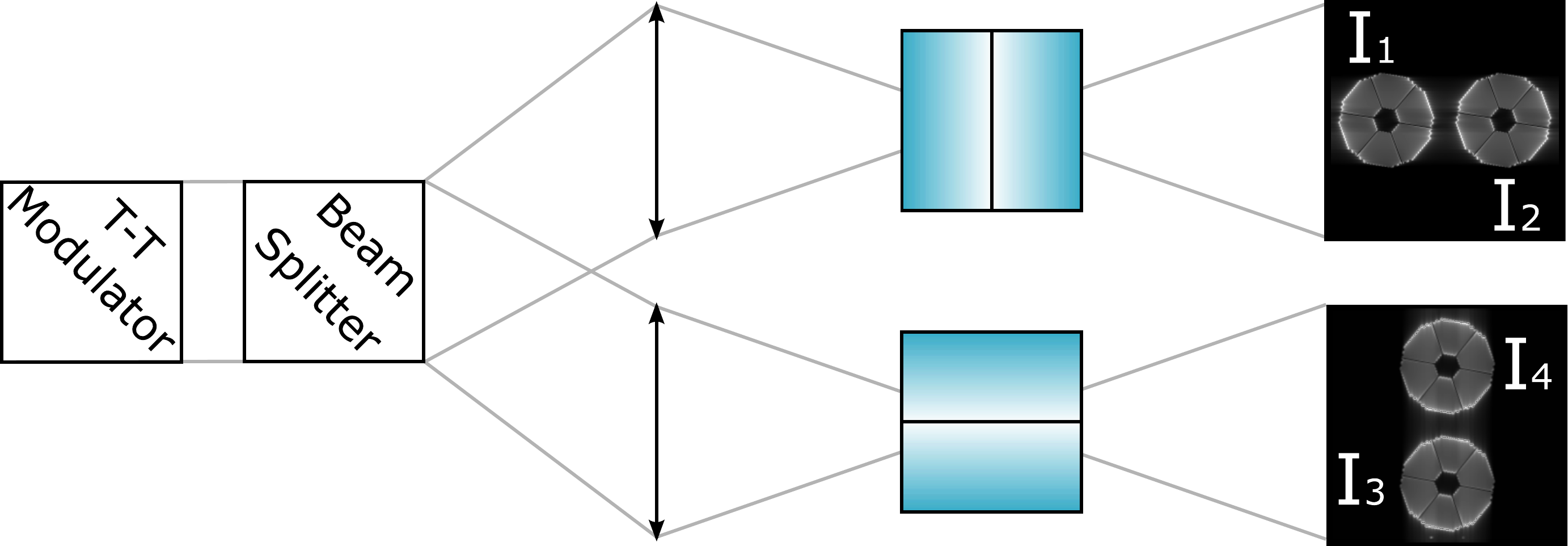} 
\caption{Schematic view of the concept of modulated Bi-O-edge sensor based on two  refractive roof prisms.}
\label{fig:BIO_EDGE_MOD_PRISM}
\end{figure}

The first new concept, named Sharp Bi-O-edge, is presented in Fig. \ref{fig:BIO_EDGE_MOD_PRISM}. It consists of a tip-tilt beam  modulator (e.g. same circular shape and amplitude than the PWS \footnote{other tip-tilt modulation schemes are possible (e. g. one per channel, uniform and linear) but are not considered here for simplicity}) followed by a $50/50$ beam-splitter. In each channel, the prism is equivalent to two genuine FKEs sharing the same edge. The respective edges in both channels are orthogonal to each other. The sensing is done by recording the intensity in the 4 re-imaged pupils. 
The prism of the Sharp Bi-O-edge has the advantage of being very easy to manufacture removing the requirement of producing a pointy tip where the sides of the Pyramid meet.

 The equivalence to FKEs is ensured at the condition that the prisms have sufficiently large deflection angle avoiding significant leakage between the two diffracted beams. This property is easily met when tip-tilt modulation is used, therefore we neglect the leakage term and consider the analysis of the sensors using pure amplitude masks here after.
 
\subsection{The Grey Bi-O-Edge}
\label{section_grey_bio}
\subsubsection{\correction{Concept}}
\label{section_grey_bio_concept}
\begin{figure}[h]
\centering\includegraphics[width=0.35\textwidth]{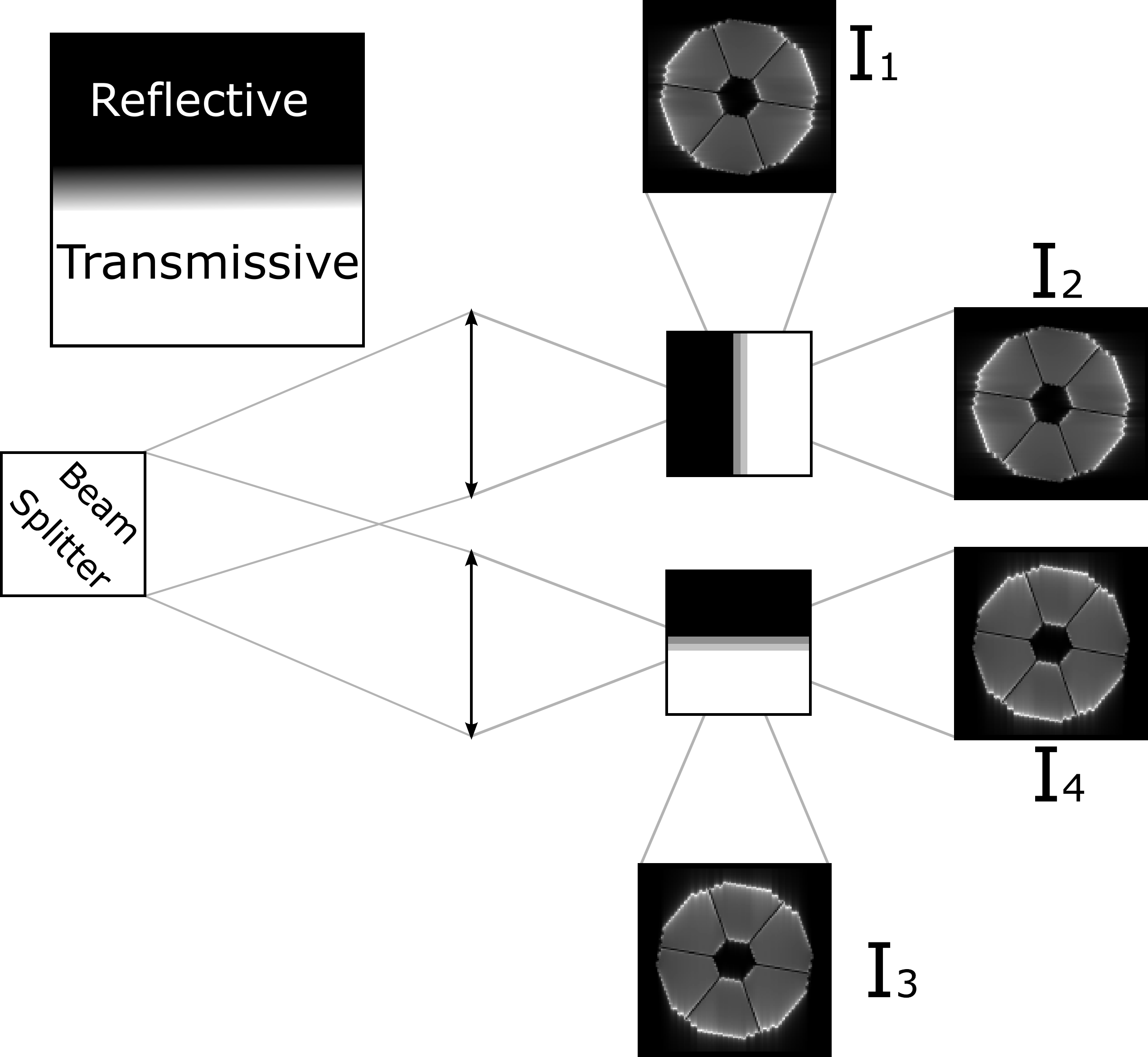} 
\caption{Schematic view of the concept of Grey Bi-O-edge with reflective/
transmissive plates. Black indicates 'reflective' and white is 'transmissive'. The grey colour denotes a gradient-like semi-reflective zone that reaches a 50\%/50\% ratio in the centre.}
\label{fig:CONCEPT_BIO_EDGE_REFL}
\end{figure}

The second new concept is a variation of the Sharp Bi-O-edge and is called Grey Bi-O-edge. It is presented in Figure \ref{fig:CONCEPT_BIO_EDGE_REFL}. In this case, the refractive facets of the prisms are replaced by a 100\% reflective or 100\% transmissive plates. The tip-tilt modulator is removed and a semi-reflective rectangular zone is present at the location of the edge of the masks that linearly goes from 100\% reflective to 100\% transmissive with the centre of the mask being exactly 50\% reflective and 50\% transmissive. This small grey gradient zone is the challenging part of the component since its width is typically of the order of the size of the modulated beam diameter, so 10s to 100s of microns in width, and shall be loss-less. 

\begin{figure}[h]
\centering\includegraphics[width=0.9\linewidth]{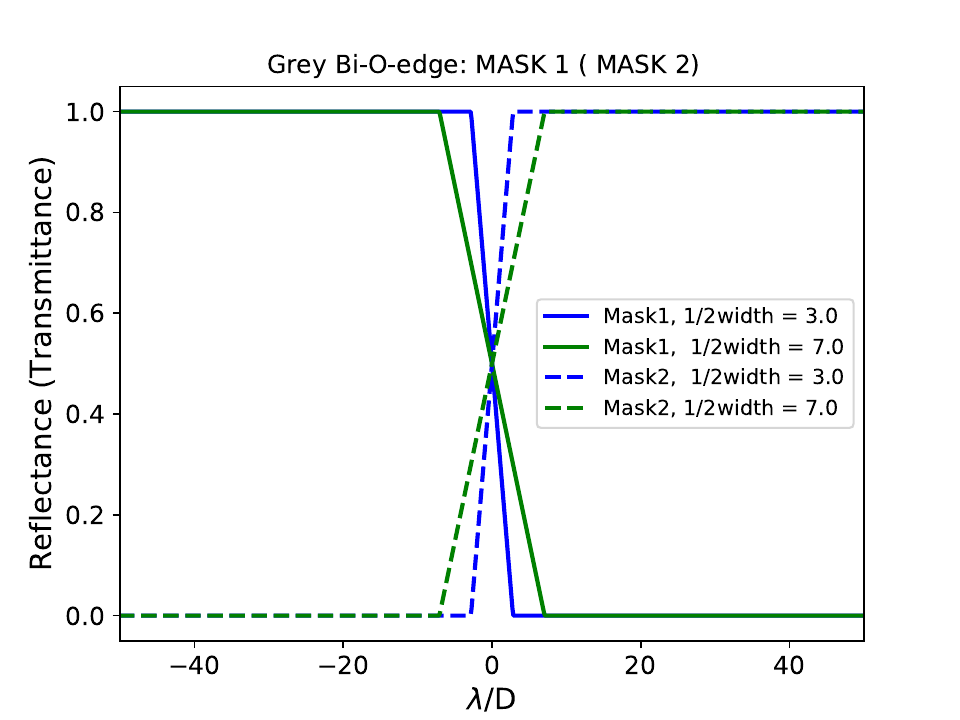} 
\caption{Reflectance (solid line) and Transmittance (dashed line) of masks 1 and 2 (modifications of $m_{bio,1}$ and $m_{bio,2}$ as defined in the appendix) of the Grey Bi-O-edge. \correction{The curves are represented for two values of the half-width of the grey zone.}}
\label{fig:BIO_EDGE_MASK_CUT}
\end{figure}

The Grey Bi-O-edge concept can also be seen as an evolution of the pupil-plane WF gradient sensor of  \cite{Horwitz_1994SPIE.2201_496H}. Indeed the Grey Bi-O-edge masks are Foucault-knife edge masks with edges having the same properties as the masks derived by \citeauthor{Horwitz_1994SPIE.2201_496H}. The latter has shown that orthogonal amplitude filters linear in intensity are equivalent to a slope sensor and can be made loss-less and symmetric. An illustration of the resulting mask (horizontal case) is provided in the top-left corner of Fig. \ref{fig:CONCEPT_BIO_EDGE_REFL} and a cut of reflectivity / transmission is represented in Fig. \ref{fig:BIO_EDGE_MASK_CUT}.

\subsubsection{\correction{Static modulation}}
\label{section_grey_bio_static_modulation}
\correction{The grey edge plays a similar role than the tip-tilt modulation: 
it reduces the sensitivity of G-modes and increases the dynamic range.
The geometrical model of \cite{Ragazzoni_Pyramid_1996JMOp_43_289R} can be adapted to the Grey Bi-O-edge: a ray originating from the pupil with some angle (local WF derivative) will be affected to 2 values, one in each quadrant: the reflectance and the transmittance at the location where the ray hits the mask. 
This way, the effect on the tip-tilt mode dynamic range can be easily understood since the signal is expected to be linear over approximately the width of the grey zone. 
For the other modes, likewise the PWS, non-linear diffraction effects are more prominent and more advanced  models are needed like in \cite{FAUVARQUE_OPTICS_2016Optic_3.1440F} or using an end-to-end model. We expect qualitatively a similar behaviour where the dynamic range is affected in an opposite way to the sensitivity and with a similar frequency dependence as suggested by the sensitivity analysis later in this paragraph.  A thorough study of the dynamic range as function of modes must definitely be part of a forthcoming study. However, we can reveal some diffraction aspects of the grey edge effect on the sensitivity in the small phase regime. } 

\correction{\cite{Guyon_XAO_2005ApJ_629_592G} determines the sensitivity of the PWS for a given Fourier G-mode by considering two configurations (see Fig. 5 from \cite{Guyon_XAO_2005ApJ_629_592G}):
\begin{enumerate}
     \item signals with maximum fringe contrast (configuration as in  Sect.\ref{sec:NATURE_FKE} when the core interferes with only one speckle)
    \item signals with a blank pupil (no signal), in a configuration where the two speckles and the core interfere.
\end{enumerate}
The total signal is obtained from a sum of signals weighted by the time passed on each configuration.}
\par
\begin{figure}[h]
\centering\includegraphics[width=\linewidth]{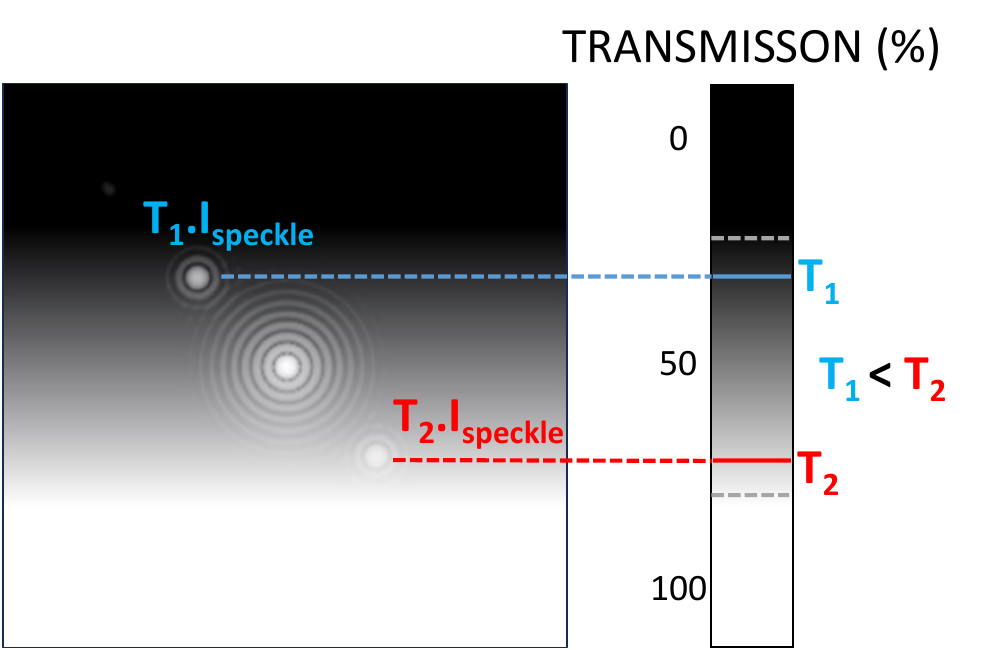} 
\caption{\correction{Static modulation mechanism in the case of the Grey edge transmitted beam. The image shows the superposition of the square of the amplitude filter with a single Fourier component PSF (G-mode). The grey edge width is indicated by two grey dashed lines on the grey-scale bar. The blue (speckle 1) and red (speckle 2) expressions indicate the intensity of the speckles after application of the amplitude mask. The signal in the pupil results from the interference of the core with two speckles with unbalanced intensity.}}
\label{fig:STATIC_MODULATION_EMPIRICAL}
\end{figure}

\correction{The mechanism of static modulation of the Grey edge is illustrated in Fig. \ref{fig:STATIC_MODULATION_EMPIRICAL} representing the superposition in transparency of a PSF and the amplitude filter (only the transmitted part is represented and the grey zone width has been exaggerated for illustrative purposes). In Section \ref{sec:NATURE_FKE}, we mentioned that the two speckles complex amplitude have a $\pi$ phase shift that leads to fringes with opposite sinusoids when each speckle interferes independently with the core. The grey edge modulation mechanism consists in a 3-source interference where the intensity of the two speckles are unbalanced by the amplitude mask. In the case presented in Fig. \ref{fig:STATIC_MODULATION_EMPIRICAL}, the signal shall have the fringes with phase obtained from the interference with speckle 2, with the coherent addition of the opposite phase fringed pattern of the interference with speckle 1. This results in a contrast damping that explains qualitatively the reduction of the signal. A quantitative evaluation with this empirical model is complex especially because of the coherent nature of the sum. Moreover, in the more practical case of a small grey width, the speckle pinning effect with the core must be taken into account. This can be done by using an end-to-end model which will also yield the complete study of the dynamic range. }

\begin{figure}[h]
\centering\includegraphics[width=\linewidth]{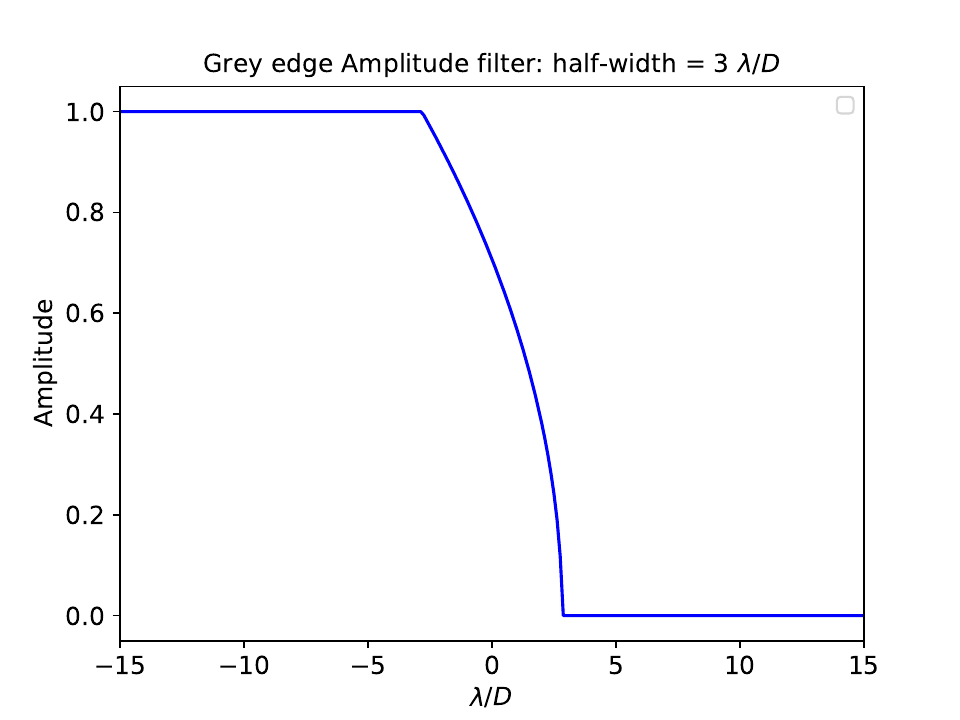} 
\caption{\correction{Amplitude filter corresponding to mask 1 of the Grey Bi-O-edge (cut along X).}}
\label{fig:STATIC_MODULATION_MASK}
\end{figure}

\correction{It is however interesting to check if the signals for G-modes are of derivative nature as shown by \cite{Horwitz_1994SPIE.2201_496H} for a pure gradient mask. Let us have a glance at the Grey edge sensitivity with the help of the C-model as in Sect. \ref{sec:FKE_PYRAMID}. We use again Eq. \ref{eq:DEF_TF_MASK} but with $r_{mod}=0$ in Eq. \ref{eq:DEF_O} and with the grey edge amplitude mask equal to the square-root of the transmittance (see Fig. \ref{fig:STATIC_MODULATION_MASK}).}

\begin{figure}[h]
\centering\includegraphics[width=\linewidth]{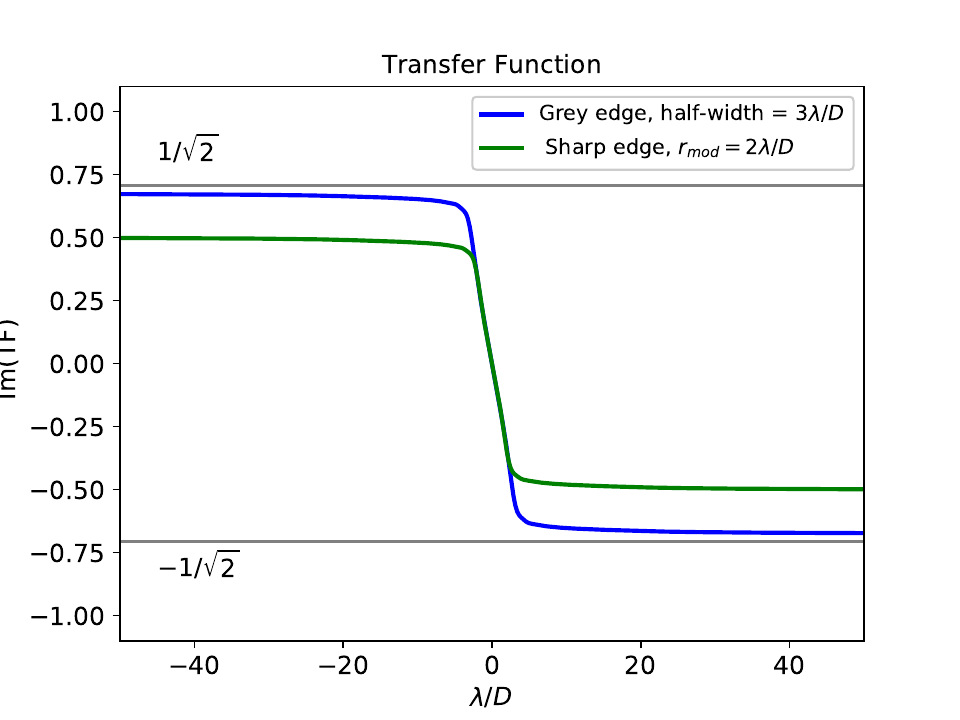} 
\caption{\correction{Imaginary part of the Transfer Function for a single quadrant (cut along X). Comparison between the Grey Bi-O-edge and the tip-tilt modulated Sharp Bi-O-edge. The modulus of the TF is linear with the spatial frequency for G-modes for both concepts which is equivalent to slope sensing. The H-modes sensitivity of the Grey Bi-O-edge is $\sqrt{2}$ larger than the one of the Sharp Bi-O-edge.}}

\label{fig:STATIC_MODULATION_TF}
\end{figure}

\correction{The transfer function (purely imaginary) for one quadrant is represented in Fig. \ref{fig:STATIC_MODULATION_TF}.
The TF of a Sharp Bi-O edge (modulation radius $2 \lambda/D$) and of a grey-Bi-O Edge (grey zone half-width $3 \lambda/D$) are provided. The grey width is slightly larger than the TT modulation to include the sensitivity damping due to the circularity of the TT modulation. This Figure shows that the grey Bi-O-edge measurements for G-modes is qualitatively similar to the one provided by the modulated Bi-O-edge, i.e. to the derivative of the phase.}

\correction{At last, as done in section \ref{sec:FKE_PYRAMID}, we can evaluate the level of the plateau of the TFs for H-modes: the Grey edge sensitivity is $\sqrt{2}$ larger than the one of the modulated Sharp edge and will also find an explanation in Sect. \ref{sec:sensitivity_noise_prop}. } 

\section{Empirical model for Sensitivity and Noise Propagation}
\label{sec:sensitivity_noise_prop}
The goal of this section is to predict the performance of the different concepts in terms of sensitivity and noise propagation for G and H-modes by observing the signal formation. We use the end-to-end model OOPAO (\cite{Heritier_OOPAO}, \url{https://github.com/cheritier/OOPAO}) and the results of Sect. \ref{sec:NATURE_FKE} to break down the signal generation along the modulation path. The case of a modulated PWS is considered as our reference and is compared to the Sharp and Grey Bi-O-edge concepts. \\
The error budget of an AO system giving the residual phase variance $\Lsigma_\phi ^2$ can be written as the sum of the fitting, temporal, aliasing and noise propagation error variances:

\begin{equation}
    \Lsigma_\phi ^2 = \Lsigma_{fitting} ^2 + \Lsigma_{temporal} ^2 + \Lsigma_{aliasing} ^2 +\Lsigma_{noise} ^2
\end{equation}
In this paper, we assume that the WFS detector is only affected by photon noise and so does the noise propagation: 
\begin{equation}
    \Lsigma_{noise}^2 = \Lsigma_{ph}^2 
\end{equation}
The derivation of $\Lsigma_{ph}$ for a FF-WFS has been given in \cite{FAUVARQUE_OPTICS_2016Optic_3.1440F}. We write this term under the assumption of a uniform pupil illumination and conservation of incident flux in the geometrical pupils. These suppositions are met in the small phase regime and if neglecting the diffraction by the edges of the masks.
$\Lsigma_{ph}$ can then be written:
\begin{equation}
    {\Lsigma}_{ph}^2 = \frac{1}{{\s(\phi)}^2}\sigma_N^2
    \label{sigma_noise}
\end{equation}
where the sensitivity $\s(\phi)$ with respect to the phase $\phi$  is defined in Eq. \ref{eq:SENSITIVITY_DEF}. In case of multiple components (like in the case of slope measurements), the sensitivity is the quadratic sum of the sensitivity for each component:
\begin{equation}
   \s(\phi)^2=\s_{x}(\phi)^2+\s_{y}(\phi)^2
    \label{eq:quadratic}
\end{equation}
$\sigma_N^2$ is the measurement variance due to photon noise only.
In the following, we use these formulas to derive a theoretical performance comparison when $\phi$ is either a G-mode or an H-mode.

\subsection{Application to the PWS}
\label{sec:sensitivy_prop_pyr}

The PWS (see Fig. \ref{fig:PYR_MOD}) produces 4 pupil images on the WFS detector. The sub-aperture resolution is then given by the WFS detector pixels sampling the pupil. For each of the 4 pupils, the image intensities are given by $I_i(x,y)$ with i being the pupil index and (x,y) being the pixel coordinates. With $N$ the incident flux per sub-aperture ($4$ pixels) and per frame, the slope-like measurement definition $S_x$ and $S_y$ with global normalisation \citep{Verinaud_Pyramid_2004OptCo.233_27V} is defined as: 
\begin{equation}
\label{eq:I_pyr_x}
\ma{S}^{pyr}_{x}(x,y) =\frac{(\ma{I}_2 + \ma{I}_4 )- (\ma{I}_1 + \ma{I}_3)}{N}
\end{equation}

\begin{equation}
\label{eq:I_pyr_y}
\ma{S}^{pyr}_{y}(x,y) =\frac{(\ma{I}_1 + \ma{I}_2 )- (\ma{I}_3 + \ma{I}_4)}{N}
\end{equation}
 In these equations, $N$ is a fixed normalisation factor while $\ma{I}_i(x,y)$ is the variable signal. The variance of the signal in each pixel is (Poisson's statistics):
 \begin{equation}
 \sigma_{I_i(x,y)}^2 = \frac{N}{4}
 \label{eq:sigma_I_i}
 \end{equation}
 
 \newcommand{\Var}{\operatorname{Var}}
 
 From Eqs. \ref{eq:I_pyr_x} and  \ref{eq:I_pyr_y} we compute the measurement noise variance  $\sigma_{N,pyr}^2 = \Var(\ma{S}^{pyr}_{x}) =\Var(\ma{S}^{pyr}_{y}) $:
\begin{equation}
\label{eq:SIGMA_MEAS_PYR}
    \sigma_{N,pyr}^2 = \frac{ \sigma_{I_1}^2 + \sigma_{I_2}^2 + \sigma_{I_3}^2 + \sigma_{I_4}^2}{N^2} = \frac{4{\frac{N}{4}}}{N^2} = \frac{1}{N}
\end{equation}

To derive the sensitivity term $\s$, it is necessary to carry out a more complex analysis: $\phi$ must be distinguished between G and H-modes and the impact of the tip-tilt modulation path must be included. We introduce the concept of modulation duty-cycle per quadrant $\ma{DC}_{mod}$ to quantify the fraction of the time during which the signal is created on a quadrant along one full modulation cycle.

For G-Modes, hence producing intensity perturbations close to the core of the PSF, the tip-tilt modulation dispatches the light in the four quadrants so that each quadrant contributes to the signal.
In these conditions, G-modes are well described by a geometric model like in \cite{Ragazzoni_Pyramid_1996JMOp_43_289R} where the measurement is a Quad-Cell-like (denoted QC) derivative and is given by Eq. \ref{eq:DERIV_FKE}. Let  $\sQ(u,v)$ denote the PWS sensitivity to the G-modes. For G-modes the sensitivity term depends explicitly on the variables $(u,v)$. However, for the sake of simplicity, we hide these variables.
Eq. \ref{sigma_noise} can then be written for the G-modes: 
\begin{equation}
        {\Lsigma_{\!\!ph,G}^{pyr}}^{\!2} = \frac{1}
        {\left(\s_{\ma{G}}^{\ma{4Q}}\right)^2}\times\frac{1}{N} 
        \label{eq:NOISE_G_MODES_PYR}
\end{equation}

For H-modes, the situation is different:  the signal is created only when the interfering core and  speckle are located in the same quadrant. We know from Eq.\ref{eq:STD_I_fringe}, that the corresponding sensitivity  is equal to unity ($\s^{CS}=1$).

\begin{figure*}[h]
    \centering
    \begin{tabular}{c c c c|c}
    1/4 frame \#1 &1/4 frame \#2&1/4 frame \#3&1/4 frame \#4& One frame\\
    \includegraphics[width=0.17\textwidth]{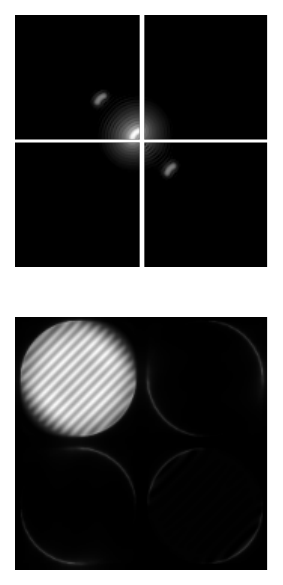}&
    \includegraphics[width=0.17\textwidth]{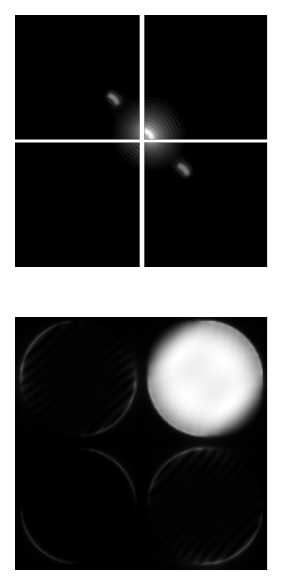}&
    \includegraphics[width=0.17\textwidth]{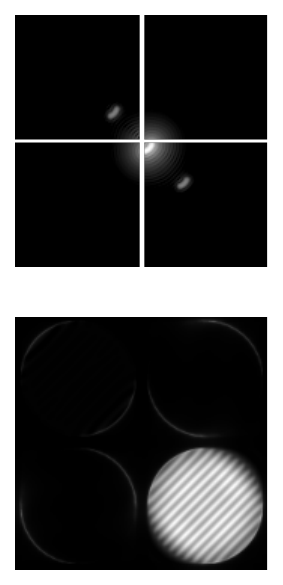}&
    \includegraphics[width=0.17\textwidth]{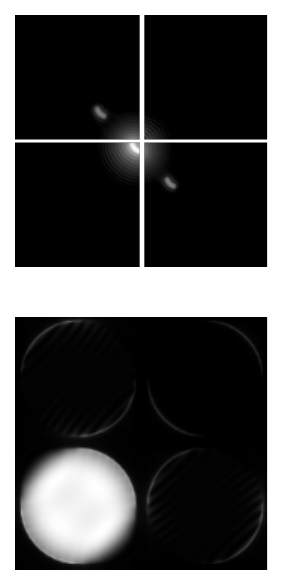}&
    \includegraphics[width=0.17\textwidth]{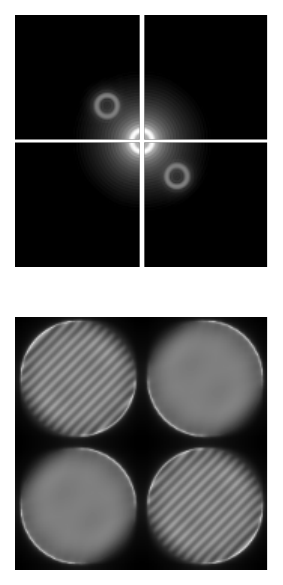}

    \end{tabular} 
    \caption{Top: View on the focal plane of the PWS during the different phases of the modulation cycle. Bottom : Corresponding signal created on the detector. The corresponding integrated modulation path and signal is displayed in the right part of the figure.   }
    \label{fig:duty_cycle_pyr}
\end{figure*}

Figure \ref{fig:duty_cycle_pyr} gives the details for the PWS and illustrates when the signal is created for the four 1/4 frame along one modulation cycle. Because of the separation between speckle and core and the shape of the masks, the signal is created only during 1/4 frames for each quadrant and two quadrants do not get any signal.  So the duty-cycle $\ma{DC}_{H}^{pyr}$ per quadrant is $25\%$.
By evaluating Eq. \ref{eq:I_pyr_x} and \ref{eq:I_pyr_y}  we get:
\begin{equation}
    \s_{H,x}^{pyr} = \s_{H,y}^{pyr} = 2 \times \s^{CS} \times \ma{DC}_{H}^{pyr} = 0.5
\end{equation}
where the factor of $2$ arises from the subtraction of the two fringed pupils $\pi$-shifted one to the other. In the example for $\s_{H,x}^{pyr}$, the signal comes from the subtraction of pupil $\#1$ from $\#3$.
Then we have:
\begin{equation}
    \left({\s_{H}^{pyr}}\right)^2 = \left({\s_{H,x}^{pyr}}\right)^2+\left({\s_{H,y}^{pyr}}\right)^2 =0.25+0.25= 0.5 
\end{equation}
Eq. \ref{sigma_noise} can then be written for the H-modes: 
\begin{equation}
        {\Lsigma_{\!\!ph,H}^{pyr}}^{\!2} = \frac{1}{0.5} \times \frac{1}{N} =\frac{2}{N} 
\end{equation}
\subsection{Application to the sharp Bi-O-edge}
\label{sec:sensitivy_prop_sharp}
For the Bi-O-edge (sharp and grey), we use the subscript 'bio' whenever an assertion is applicable to both the sharp ('sha') and the grey ('gre') concept.  We define the measurements in X and Y as:
\begin{equation}
\label{eq:I_bio_x}
\ma{S}^{bio}_{x}(x,y) = \frac{\ma{I}_2 - \ma{I}_1 }{N/2}
\end{equation}
\begin{equation}
\label{eq:I_bio_y}
\ma{S}^{bio}_{y}(x,y) = \frac{\ma{I}_4 - \ma{I}_3 }{N/2}
\end{equation}
This definition takes into account that the flux is split into two channels, each receiving half of the flux.  The corresponding measurements variance due to photon noise is: 
\begin{equation}
    \sigma_{N,bio}^2 = \frac{ \sigma_{I_1}^2 + \sigma_{I_2}^2}{(N/2)^2} =\frac{\sigma_{I_3}^2 + \sigma_{I_4}^2}{(N/2)^2} = \frac{2{\frac{N}{4}}}{(N/2)^2} = \frac{2}{N}
    \label{eq:sigma_phot_BIO}
\end{equation}
We can notice that the Bi-O-edge measurement variance $\sigma_{N,bio}^2$ is twice the one of the PWS given by Eq.\ref{eq:SIGMA_MEAS_PYR}.

For G-modes, the geometrical model used in \cite{Ragazzoni_Pyramid_1996JMOp_43_289R} can be directly applied to the Sharp Bi-O-edge \footnote {the sum of two PWS pupils being equivalent to an FKE (in the geometric approximation) as mentioned by \cite{Ragazzoni_Pyramid_1996JMOp_43_289R}} and gives the same result:
\begin{equation}
    \s_{G}^{sha} = {\sQ}
\end{equation}
The noise propagated on G-modes for the sharp Bi-O-edge is:
\begin{equation}
        {\Lsigma_{ph,G}^{sha}}^{\!\!2} = \frac{1}{\left({\sQ}\right)^2} \times \frac{2}{N} 
        \label{eq:NOISE_G_MODES_BIO}
\end{equation}

The G-mode noise propagation of the sharp Bi-O-edge is therefore twice higher than for the PWS. This behaviour was intuitively expected as X- and Y-slopes are only derived using half of the total flux because of the beam-splitting. For the PWS instead, the slopes are calculated using all available photons.

\begin{figure*}[h]
    \centering
    \begin{tabular}{c c c c|c}
    1/4 frame \#1 &1/4 frame \#2&1/4 frame \#3&1/4 frame \#4& One frame\\
    \includegraphics[width=0.17\textwidth]{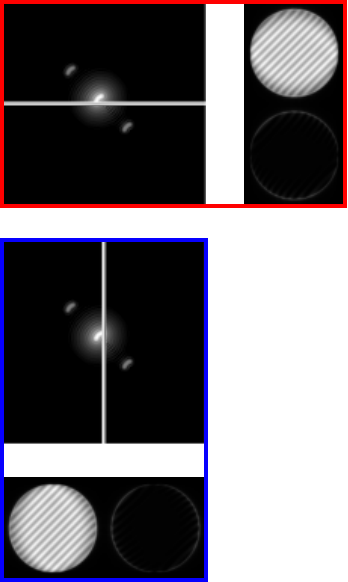}&
    \includegraphics[width=0.17\textwidth]{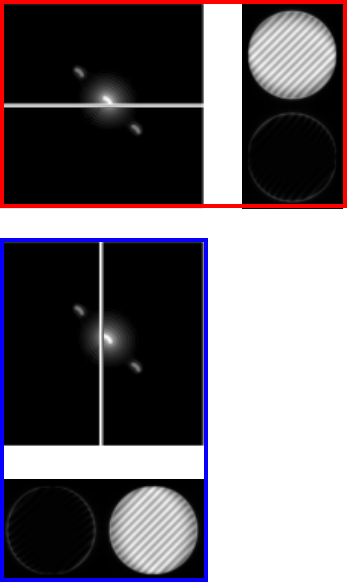}&
    \includegraphics[width=0.17\textwidth]{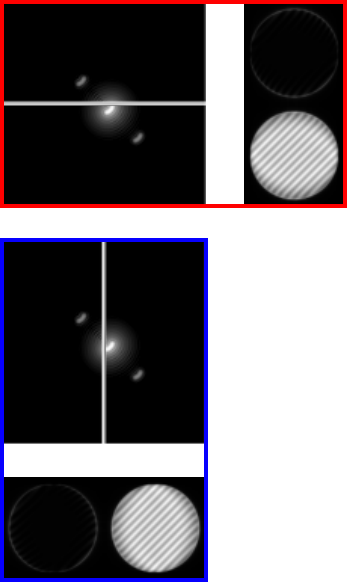}&
    \includegraphics[width=0.17\textwidth]{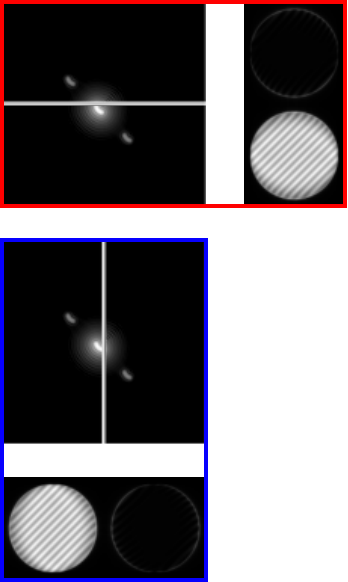}&
    \includegraphics[width=0.17\textwidth]{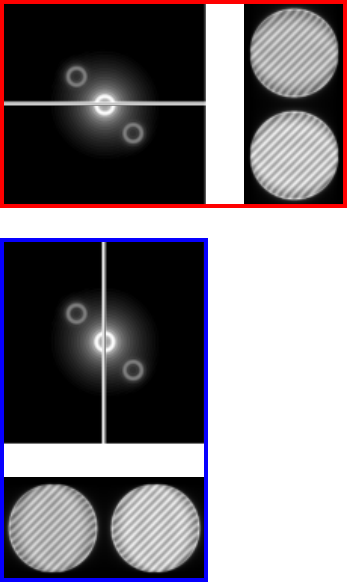}

    \end{tabular}
    \caption{Signal created for each quarter of a full modulation cycle for the Bi-O-edge . The red frames correspond to the first channel(horizontal split) and the blue frames to the second channel (vertical split). }
    \label{fig:duty_cycle_bio}
\end{figure*}

For the H-modes, the details of the modulation cycle for a Sharp Bi-O-edge are provided in Figure \ref{fig:duty_cycle_bio} and shows that contrarily to the PWS , all quadrants provide a signal 50\% of the time ($\ma{DC}_{H}^{sha} = 0.5$)

We can compute then the associated sensitivity of the Sharp Bi-O-edge as: 
\begin{equation}
    \s_{H,x}^{sha} = \s_{H,y}^{sha} = 2 \times \s^{CS} \times \ma{DC}_{H}^{sha} = 1
\end{equation}
The complete sensitivity term is:
\begin{equation}
    \left({\s_{H}^{sha}}\right)^2 = \left({\s_{H,x}^{sha}}\right)^2+\left({\s_{H,y}^{sha}}\right)^2 =1+1= 2
\end{equation}
Eq. \ref{sigma_noise} can then be written for the H-modes: 
\begin{equation}
        {\Lsigma_{\!\!ph,H}^{sha}}^{\!2} = \frac{1}{2} \times \frac{2}{N} =\frac{1}{N} 
\end{equation}
The H-mode noise propagation of the sharp Bi-O-edge is therefore twice lower than for the PWS. Also, this behaviour can be understood intuitively: The Bi-O-edge generates signal all the time during a modulation cycle while the PWS is blind to a particular H-mode half of the time (e.g., during 1/4 frames \#2 and \#4 in Fig. \ref{fig:duty_cycle_pyr}).

\subsection{Application to the Grey Bi-O-edge}
\label{sec:sensitivy_prop_grey}

\begin{figure}[h]
    \centering
    \includegraphics[width=0.25\textwidth]{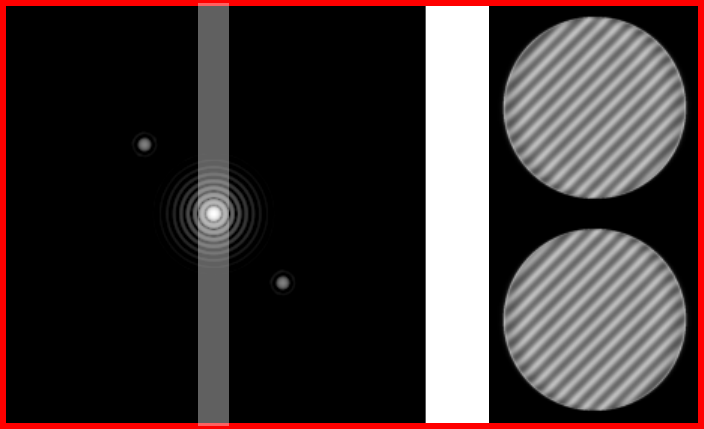} 
    \hspace{0.25cm}
    \includegraphics[width=0.15\textwidth]{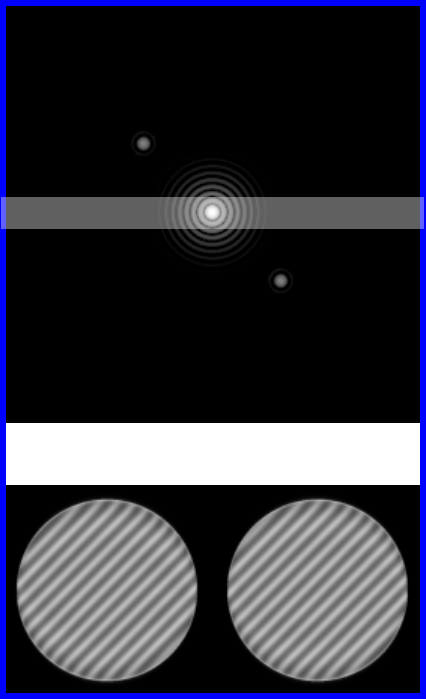}
    \caption{Signal creation for the Grey Bi-O-edge. The grey stripe represents the zone with gradient-shape reflectivity/transmissivity.}
    \label{fig:SIGNAL_BIO_GREY}
\end{figure}

We assume that the width of the grey zone is  $\pi/2$ times larger than the diameter of the circular modulation in order to account for the difference in sensitivity between linear and circular shape. Under these assumptions, and since the measurements definition is the same as in Eqs. \ref{eq:I_bio_x} and \ref{eq:I_bio_y}, we have:
\begin{equation}
    {\s_{G}^{gre}} ={\sQ}
\end{equation}

\begin{equation}
        {\Lsigma_{ph,G}^{gre}}^{\!\!2} = \frac{1}{\left({\sQ}\right)^2} \times \frac{2}{N} 
        \label{eq:NOISE_G_MODES_GRE}
\end{equation}

The Grey Bi-O-edge signal formation for H-modes is represented in Fig.\ref{fig:SIGNAL_BIO_GREY}. As for the Sharp Bi-O-edge there is no blind zone. The efficiency is $100\%$ duty cycle. However, since the flux in the core is split equally between reflected and transmitted beams, the sensitivity $\s^{\ma{1/2CS}}$ corresponding to the interference between half of the core and a speckle is reduced by a factor $\sqrt{2}$ in accordance with Eq. \ref{eq:I_CORE_SPECKLE} and we have:
\begin{equation}
\label{eq:1s2CS}
    \s^\ma{1/2CS} = \frac{1}{\sqrt{2}} 
\end{equation}
We can then compute the sensitivity of the Grey Bi-O-edge as: 
\begin{equation}
    \s_{H,x}^{gre} = \s_{H,y}^{gre} = 2 \times \s^\ma{1/2CS} = \sqrt{2}
\end{equation}
The complete sensitivity term is:
\begin{equation}
\left({\s_{H}^{gre}}\right)^2 = \left({\s_{H,x}^{gre}}\right)^2 + \left({\s_{H,y}^{gre}}\right)^2=2+2= 4
\end{equation}

Eq. \ref{sigma_noise} can then be written for the H-modes: 
\begin{equation}
        {\Lsigma_{\!\!ph,H}^{gre}}^{\!2} = \frac{1}{4} \times \frac{2}{N} =\frac{1}{2N}  
\end{equation}

The H-mode noise propagation of the grey Bi-O-edge is therefore twice lower than for the sharp Bi-O-edge and 4 times lower than for the PWS. This represents a significant advantage for AO which is frequently struggling with the limited number of photons provided by the AO guide star. Intuitively, the Grey Bi-O-edge makes better use of the photons because H-modes produce signal in all quadrants all the time instead of only half of the time for the Sharp Bi-O-edge with modulation. Even considering the fringe contrast loss of $\sqrt{2}$ (Eq.\ref{eq:1s2CS}), this leads to a net gain of a factor of $2$ in efficiency to the use of photons. This is analogous to the improved sensitivity of the non-modulated PWS over the modulated one even for very small modulation \citep{Guyon_XAO_2005ApJ_629_592G}.

\subsection{Summary}
\label{noise_propag_summary}

We summarise in table \ref{tab:summary} all the findings of Sect.\ref{sec:sensitivity_noise_prop}. The result of Eq. \ref{sigma_noise} represents the behaviour of the noise propagation for the different concepts for G and H-modes. 

For G-modes, a PWS presents twice less noise propagation ($1/N$) than the Bi-O-edge concepts ($2/N$) because the split of light in the latter reduces by a factor of $2$ the number of photons for measuring each component of the derivative. 

For H-modes, the Sharp Bi-O-edge presents twice less noise propagation ($1/N$) than the PWS ($2/N$). One way to understand it, is that the Hilbert transform carries all the information and hence does not suffer from the split of light (in our read-noise free hypothesis). Moreover, the Fourier masks of the PWS are such that two quadrants are blind to a given H-mode. The Grey Bi-O-edge gains another factor of $2$ ($1/(2N)$) with respect to the Sharp Bi-O-edge ($1/N$) because of the static nature of the modulation.

Section \ref{sec:Modal_Analysis} analyses in details how the noise propagation is distributed on the G and H-modes and how this determines the overall noise propagation.

\begin{table}
\begin{tabular}{|c||c|c|c|}
\hline
 & & & \\
 & PWS & Sharp Bi-O-edge & Grey Bi-O-edge \\
  & & & \\
\hline\hline
 & & & \\
G-modes & {\Large$\frac{1}{N}$} &  {\Large$\frac{2}{N}$} &  {\Large$\frac{2}{N}$} \\
& & & \\
\hline
& & & \\
H-modes & {\Large $\frac{2}{N}$}&  {\Large $\frac{1}{N}$} &  {\Large$\frac{1}{2N}$} \\
& & & \\
\hline
\end{tabular}
\caption{\label{tab:summary}Empirical estimation of the noise propagation (Eq.\ref{sigma_noise}). For G-modes, the dependency on ${\sQ(u,v)}^2$ has been factored out.}
\end{table}

\section{Analysis of the Bi-O-edge performance gains}
\label{sec:Modal_Analysis}
\subsection{Modal noise propagation}
\label{sec:MODAL_NOISE_C-MODEL}
The empirical model of Section \ref{sec:sensitivity_noise_prop} makes a rigid distinction between G- and H-modes without considering how many modes of each type are actually in the system and neglecting the fact that modes with a spatial frequency around the modulation circle have mixed properties. In practice, small modulation angles (few $\lambda/D$) are used in PWS systems and the number of DoF is limited for various reasons and can be very diverse in the AO systems.  We perform in this section an improved analysis in order to derive the trend of performance gains  in function of controlled modes. 

We do the analysis for systems with realistic configurations and number of DoF and for a typical modulation radius of $2$ $\lambda/D$ (half-width=$3\lambda/D$ for the Grey Bi-O-edge). Our reference case is the ELT Single Conjugate Adaptive Optics (SCAO) first light system which is the Phasing and Diagnostic Station (PDS) PWS-based AO system with $3000$ modes controlled \citep{BONNET_AO_ELT_2018SPIE10703E_10B}. This number of modes is conservative but is found to match well the analytical model based on a least-square reconstruction developed in the appendix. 
 
We developed in appendix \ref{sec:C_model}, the formalism of the C-model of \cite{Fauvarque_Kernel_2019JOSAA_36.1241F} of the PWS and Bi-O-edge variants and derived the modal properties of  noise propagation.

\begin{figure}[h]
\centering\includegraphics[width=\linewidth]{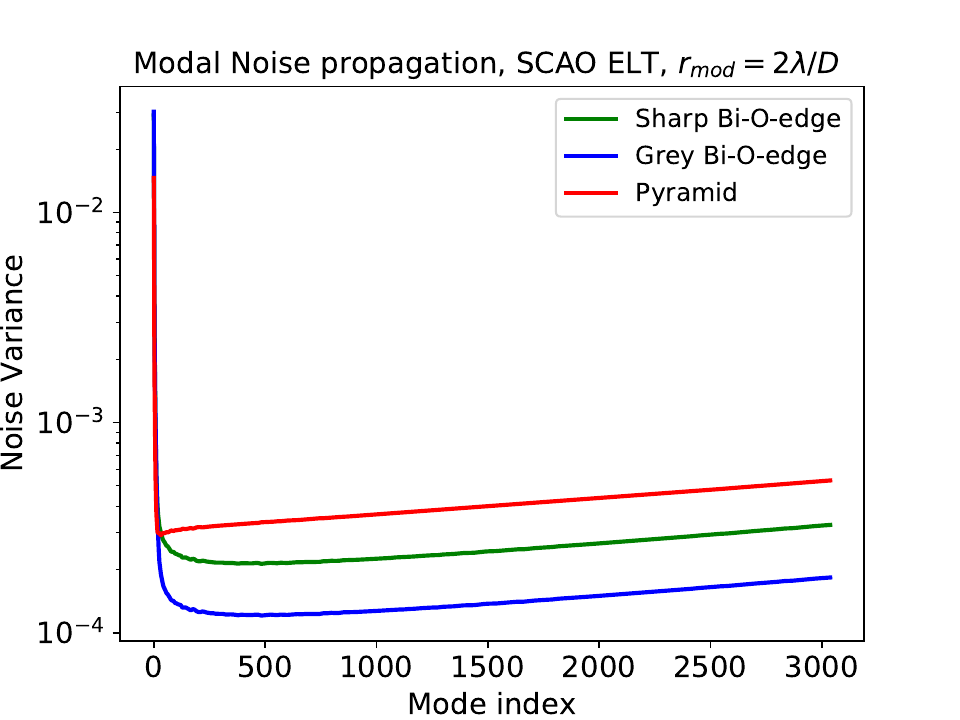} 
\caption{Noise propagation per modes for SCAO/ELT configuration (3000 modes, sensing in K band) for $2 \lambda/D$ modulation radius.}
\label{fig:NC_MODAL_C_MODEL_LIN}
\end{figure}

\begin{figure}[h]
\centering\includegraphics[width=\linewidth]{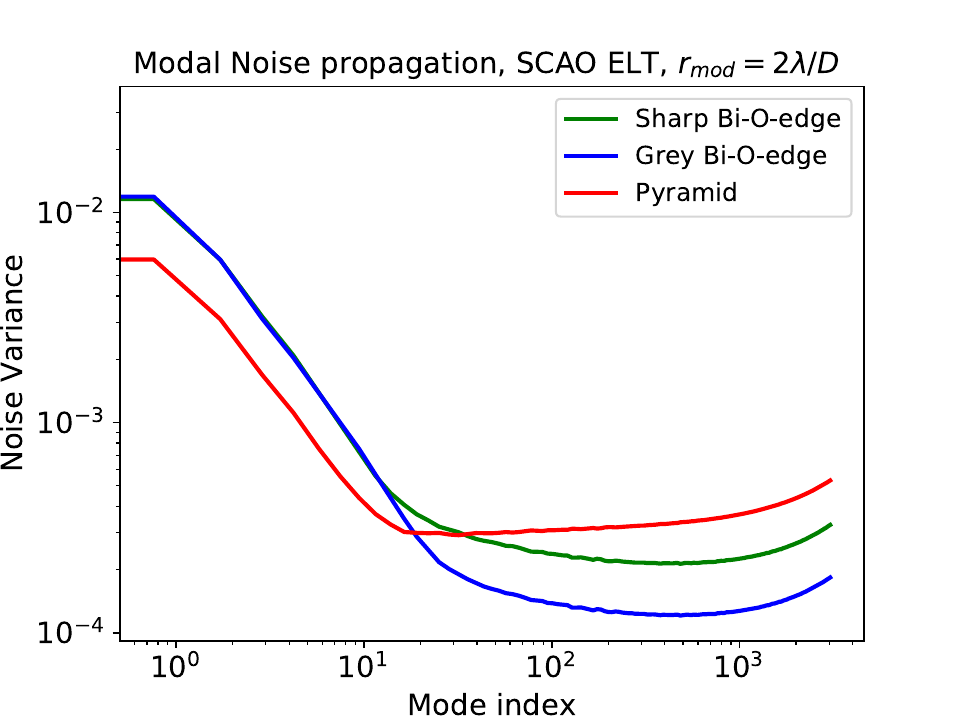} 
\caption{Same as Fig. \ref{fig:NC_MODAL_C_MODEL_LIN} with x-axis in log-scale .}
\label{fig:NC_MODAL_C_MODEL_LOG}
\end{figure}

An example of modal noise propagation curves obtained using this formalism (Eq. \ref{eq:KL_Order} and Eq. \ref{eq:NC_Fr}) are shown in Fig. \ref{fig:NC_MODAL_C_MODEL_LIN} for the $3$ concepts and for the ELT  SCAO configuration. $3000$ modes are corrected and the sensing is done in K band to stay in the linear regime which is the assumption in this paper. The figure shows that the noise propagation of the high order H-modes follows the expected tendency but the relative gains are somewhat reduced (factor $\approx 1.6$ and $\approx 3$ gain by the Grey Bi-O-edge over the Sharp Bi-O-edge and the PWFS, respectively, instead of factors of $2$ and $4$). Fig. \ref{fig:NC_MODAL_C_MODEL_LIN} also shows that even though the noise propagation on G-modes/low orders is very large, the number of H-modes is largely dominant for the chosen modulation amplitude. Fig.\ref{fig:NC_MODAL_C_MODEL_LOG} permits to see a close-up on the low order/G-modes.

The overall performance is given by the total noise propagation error variance $\ma{V}_{\ma{wfs}}$ and is obtained by integrating Eq. \ref{eq:NOISE_PROPAG_GEN}:
\begin{equation}
\label{eq:NC_WFS_integ}
\ma{V}_{\ma{wfs}} = \iint_\mathcal{A} \sigma_{\ma{wfs}}^2(u,v)du dv
\end{equation}
where the theoretical integration area $A$ is:
\begin{equation}
\label{eq:AREA_integ}
\mathcal{A}: f_{min}=\frac{1}{2D} < \sqrt{u^2+v^2} < f_{max}=\frac{1}{2d}
\end{equation}

In order to estimate the overall sensitivity gain with respect to the PWS, we define in Eq . \ref{eq:GAIN_BIO} the gain ratio $\mathcal{G}_{\ma{wfs/pyr}}(i_{kl})$ where wfs is the Sharp Bi-O-edge or the Grey Bi-O-edge. From Section \ref{sec:sensitivity_noise_prop}, we know this factor is comprised between $1/2$ (in the case of very low order system only G modes matter) and $4$ (for a grey Bi-O-edge system with very high orders dominating).  
\begin{equation}
\label{eq:GAIN_BIO}
\frac{1}{2} < \mathcal{G}_{\ma{wfs/pyr}} = \frac{\ma{V}_{\ma{pyr}}}{\ma{V}_{\ma{wfs}}} < 4
\end{equation}
The gain in function of the number of modes actually corrected can be computed by adapting the integration area in Eq. \ref{eq:NC_WFS_integ} as the cut-off frequency $f_c=1/(2d)$ is never really attained in a real system. For the SCAO ELT configuration we simulated the PDS case with $3000$ number of modes, which corresponds to a cut-off frequency of $0.7 f_c$. We use this value as the upper limit of the integration and for each value of $d$ the number of modes is adjusted accordingly. We also use a conservative approach to include the low order noise propagation: some preliminary work which goes beyond the scope of this paper,  suggests that the C-model limitations related to the integration over a finite pupil \citep{Fauvarque_Kernel_2019JOSAA_36.1241F}, leads to an underestimation of the noise propagated on low orders. We found out that extending the integration down to $f_{min}=1/(4D)$ gives an improved estimation of the low orders contribution that is sufficient to get the right trends. The corrected overall noise propagation variance $V'$ is obtained by integrating the circular averaged noise propagation for different values of $d$ in order to predict the gain for different number of corrected modes:

\begin{equation}
\label{eq:NC_WFS_integ_corrected}
\ma{V}^{'}_{\ma{wfs}} = \int_{\frac{1}{4D}}^{0.7 f_c} <\sigma_{\ma{wfs}}^2(u,v)>_{f}\cdot 2 \pi f df
\end{equation}

\begin{figure}[h]
\centering\includegraphics[width=\linewidth]{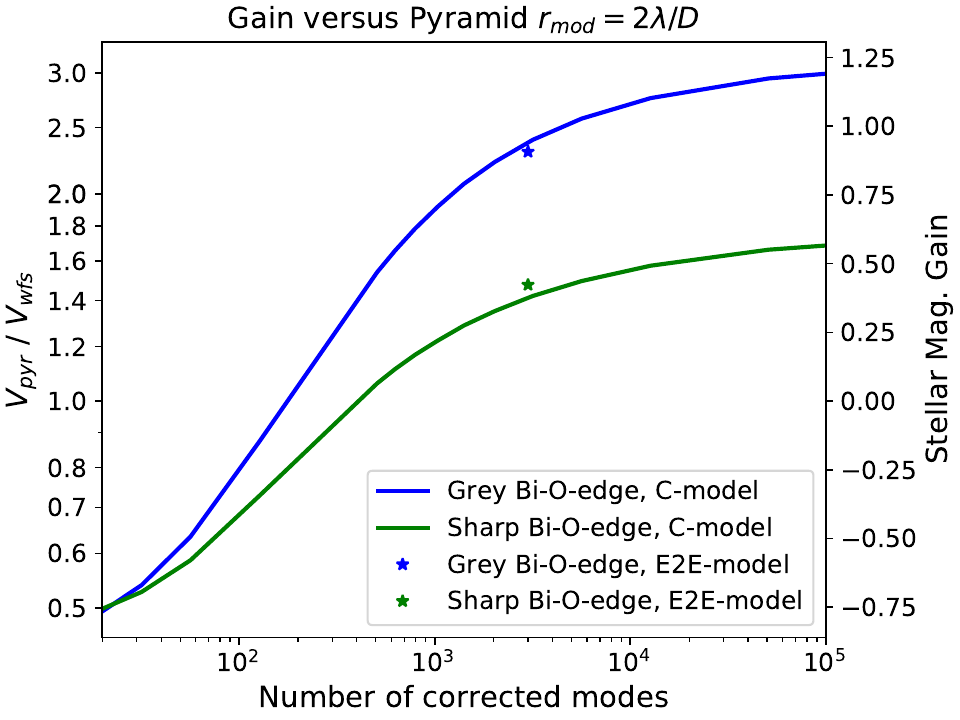} 
\caption{Gain with respect to the PWS in function of corrected modes for SCAO/ELT configuration ($3000$ modes, $2 \lambda/D$ modulation radius, sharp-Bi-O-edge and Grey Bi-O-edge).}
\label{fig:GAIN_BIO_EDGEs_vs_PYR}
\end{figure}

The result is represented in Fig. \ref{fig:GAIN_BIO_EDGEs_vs_PYR} that describes the gain in function of corrected modes. The two additional points have been computed from the noise propagation obtained using calibration data of the end-to-end model for the ELT SCAO system with 3000 modes controlled.

We can observe that for both sharp and grey Bi-O-edge concepts, the low order systems limit shows the expected loss a factor of $2$ in photon efficiency. For very high order systems ($10^5$ modes), the gain for the grey Bi-O-edge reaches a factor of $3$ ($1.2 mag$) while the sharp Bi-O-edge is limited to $1.6$ ($0.6 mag$). The validation through end-to-end closed loop simulations is presented in the next section.

\subsection{End-to-end simulations}
\label{sec:E2E_SIMULATIONS}

We developed a diffractive model of the two Bi-O-edge concepts and added it to the OOPAO package. We simulate pure amplitude masks for both the PWS and the Bi-O-edge. Our reference configuration for the simulation is the one of ELT SCAO with 3000 modes. 

\begin{figure}[h]
\centering\includegraphics[width=\linewidth]{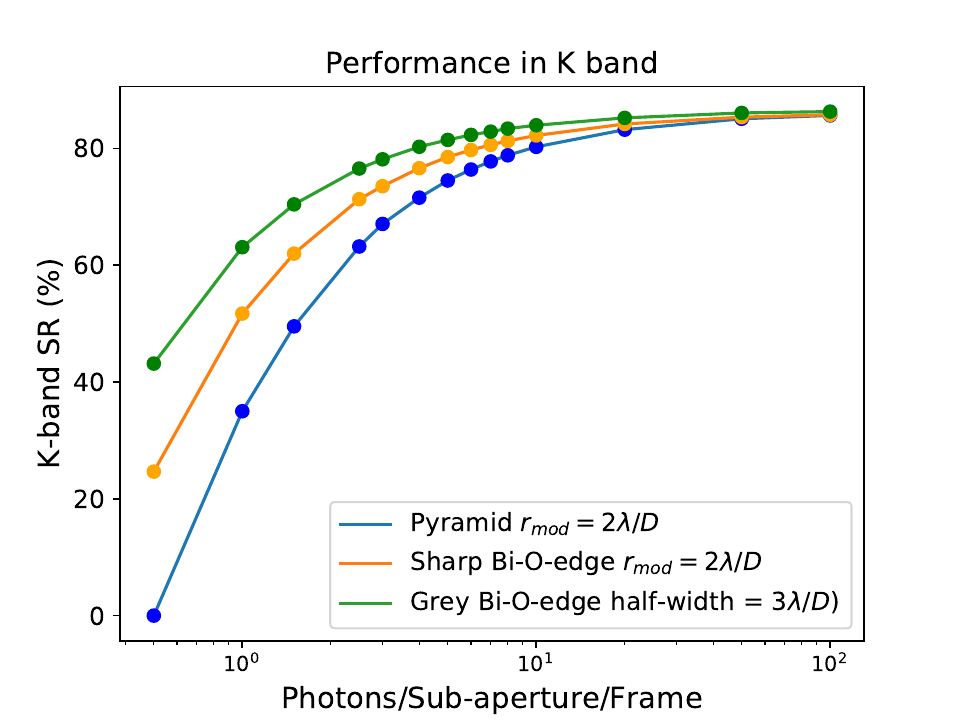} 
\caption{SR in function of flux for the SCAO/ELT configuration (3000 modes, $2 \lambda/D$ modulation radius).}
\label{fig:SR_E2E}
\end{figure}

The main parameters used in the simulation can be found in table \ref{tab:def_model}. We use a SCAO ELT K-band case where the sensors are used close to their linear regime, which is the assumption of this paper. Moreover, we suppose null read-out noise such that the performance degradation at low flux is dominated by photon noise error propagation. The performance in terms of SR and in function of the flux is displayed in Fig. \ref{fig:SR_E2E}. 

These results confirm the gains obtained with the analytical model. For instance, for a relative drop of SR of $25\%$, so for $\ma{SR}=64.5\%$, the number of photons needed are $2.7$ for the PWS, $1.75$ for the Sharp Bi-O-edge and $1.11$ for the Grey Bi-O-edge. This gives a gain versus the PWS of $1.54$ for the Sharp and $2.43$ for the Grey Bi-O-edge. Looking at Fig. \ref{fig:GAIN_BIO_EDGEs_vs_PYR}, the gains predicted from the C-model are $1.41$ and $2.33$ respectively, hence slightly pessimistic but very close to the end-to-end results.  

\correction{As a side note, we mention that for simplicity we limited the content of the paper to one modulation angle ($r_{mod} = 2 \lambda/D$ and grey edge half-width = $3\lambda/D$) that we believe is realistic for XAO with reasonably small residuals. We concentrated the effort on this case, to obtain consistent results between analytical and end-to-end simulations. We initiated some work to consider different modulation angles that showed that some aspects of the theoretical model must be adjusted. From these preliminary results, we observe that the behaviour at high number of corrected modes is merely independent of the modulation angle. We appraise that the gain for $r_{mod} = 3 \lambda/D$ would be about 5\% lower than for $r_{mod} = 2 \lambda/D$ for more than $10^4$ corrected modes. The biggest impact is on the tipping point where the Bi-O-edge gain is larger than one, i.e. when the G-modes noise propagation becomes small in the error budget. Figure \ref{fig:GAIN_BIO_EDGEs_vs_PYR} shows that more than 400 modes are needed to see a gain with the Sharp Bi-O-edge with $r_{mod} = 2 \lambda/D$. We estimate this number to be close to 100 modes for $r_{mod} = 1 \lambda/D$ and 800 modes for $r_{mod} = 3 \lambda/D$. Future work will refine this analysis and also include the Grey Bi-O-edge. We also want to mention that the case $r_{mod}=0$ was discarded because it cannot be well treated with the assumptions of this paper (uniformity and flux conservation in the pupil).}

It is also remarkable to see that for a flux of only $0.5$ photons/sub-aperture/frame, the performance of the Bi-O-edge concepts is still decent, with only $50\%$ of drop of SR for the Grey Bi-O-edge, while the PWS is not able to close the loop. These analysis will certainly need to be improved, by taking into account an optimisation of the control with respect to flux but they definitely validate the overall analysis of the Bi-O-edge concepts sensitivity advantage. \correction{Future analysis will also consider detector read-out-noise in the simulations. The factor of 2 of noise variance between the PWS and Bi-O-edge will remain when considering additional read-out-noise. This can be shown by rewriting Eqs. \ref{eq:SIGMA_MEAS_PYR} and \ref{eq:sigma_phot_BIO} where $ron$ is the read-out-noise in photo-electron per pixel. However, since the read-out-noise term has a quadratic dependence with number of photons, the Stellar Magnitude Gain in Fig. \ref{fig:GAIN_BIO_EDGEs_vs_PYR} would be reduced depending on read-out-noise but also on the stellar flux itself. Modern Avalanche-Photo-Diode-based cameras can have read-out noise as low as 0.6 electrons \citep[e.g.][]{FEAUTRIER_C-RED_ONE_2022SPIE12183E..2EF}. This will clearly affect strongly the SR curves of Fig. \ref{fig:SR_E2E} at very low flux. However, in HCI, the limitation due to stellar flux will most probably first affect the halo of residuals (the contrast) before a noticeable drop of the SR occurs. With a refined criterion relevant for HCI, the corresponding limiting magnitudes would occur at higher fluxes, such that the additional read-out-noise term would be not far from one. For instance, with $ron=0.6$ and for $N=10$ photons, the corrective term is $1.114$ while it is $2.44$ for $N=1$.}

\begin{equation}
\label{eq:SIGMA_MEAS_PYR_RON}
    \sigma_{N,ron,pyr}^2 = \frac{4{\left(\frac{N}{4}+ron^2\right)}}{N^2} = \frac{1}{N}\left(1+\frac{4ron^2}{N}\right)
\end{equation}

\begin{equation}
\label{eq:SIGMA_MEAS_BIO_RON}
    \sigma_{N,ron,bio}^2 = \frac{2{\left(\frac{N}{4}+ron^2 \right)}}{(N/2)^2} =  \frac{2}{N}\left(1+\frac{4ron^2}{N}\right)
\end{equation}

\begin{table}
    \begin{centering}
    \begin{tabular}{|c|c|c|}	
	\cline{1-3}
	\multirow{4}{*}{\textbf{Turbulence}}&Fried Parameter $r_0$ & 13 cm @500 nm \\
	&Outer Scale $L_0$ &25 m\\
	&C$n^2$ profile&1 layer\\
	&Wind-speed& 10 m/s\\
	\cline{1-3}		\multirow{4}{*}{\textbf{Control}}&Frequency&1 kHz\\
	&Integrator& $g$=0.5\\
	& Int. Matrix & 3000 KL modes \\
	\cline{1-3}	
    \multirow{3}{*}{\textbf{NGS}}&Wavelengths&K band -- 2179 nm\\
    &Photons/subap.&0.1-1000\\
    \cline{1-3}	
	\multirow{4}{*}{\textbf{Telescope}}&Diameter&39 m\\
    &Pupil&ELT-M1 pupil mask\\
    &Spiders diameter& 51 cm\\
    &Resolution&576 pix of 0.07m\\
    \cline{1-3}	
    \multirow{4}{*}{\textbf{DM}}&Actuator&5352\\
	&Geometry&Hexagonal\\
	&Inf. Functions&From ELT-M4 FEM\\
	& Coupling&0\%\\
	\cline{1-3}	
    \multirow{2}{*}{\textbf{PWFS}}&Sub-aperture size& 41.8 cm\\
	&Modulation& 2 $\lambda$/D\\
	\&&RON&None\\
	\multirow{2}{*}{\textbf{Bi-O-edge}}&Photon Noise&Yes\\
	&Signal Processing& Slopes-Maps\\
	\cline{1-3}	

	\end{tabular}
    \end{centering}
    \caption{Numerical Simulations parameters}
    \label{tab:def_model}
\end{table}

\section{Conclusion}
\label{sec:discussions}
We revisited the concept of dual channel two-sided PWS by realising that this concept is actually the implementation of two orthogonal FKEs. In order to keep the dynamic range of the PWS, we introduced the tip-tilt modulation to the concept, denoted as the sharp Bi-O-edge concept. By designing a reflective version, we realised that the modulation functionality can be achieved by implementing a reflective/transmissive central stripe with a gradually changing reflectivity and transmission. We dubbed this concept the Grey Bi-O-edge.

We used an empirical model to evaluate the efficiency of the FKE masks as used by the different concepts.
Splitting the light between two channels penalises the low orders (or G-modes), such that both Bi-O-edge flavors would actually need twice more photons than a PWS to reach the same noise propagation level in a low order system. 

However, in a high order system, the amount of noise propagated on low orders/G-modes is relatively small compared to the H-modes. These high order modes have the remarkable property that their measurement from the FKE in both channels carries the same information. This redundancy is responsible for the fact that the sensitivity is not impaired by the separation in flux and is covering the full Fourier plane. This is not the case for the PWS as shown by the TFs of the amplitude masks in Fig.\ref{fig:tf_comparison_pyr_fke} and the modulated signal decomposition in Fig.\ref{fig:duty_cycle_pyr}: each of the $4$ PWS pupils is blind to one half of the Fourier plane. This lack of the PWS also explains why the sharp Bi-O-edge exhibits, in the empirical model, a factor of 2 higher photon efficiency for H-modes. Even better, the Grey Bi-O-edge, because it has a $100\%$ modulation duty cycle, gains another factor 2 over the sharp Bi-O-edge that has only a $50\%$ duty cycle.

\correction{The empirical model of Sect. \ref{sec:sensitivity_noise_prop} is very simplified on purpose. For a study with much greater generality than ours, we advise the reader to consider \cite{CHAMBOULEYRON_NOISE_2023A&A...670A.153C}. In this paper, the notion of photon noise sensitivity has been updated and shown to have an upper limit at $2$, twice more than what was thought before \citep[eg. see][]{Guyon_XAO_2005ApJ_629_592G}. Figure 1 of  \cite{CHAMBOULEYRON_NOISE_2023A&A...670A.153C} shows how the Zernike sensor sensitivity (for H-modes) can be increased until very close to 2 at the expense of major loss of sensitivity for G-modes. Table \ref{tab:summary} of the present paper can directly be used to determine this same sensitivity by identifying the coefficient in front of $1/N$ as the inverse of the photon sensitivity squared. This gives for the photon sensitivity,  a value of $1/\sqrt{2}$ for the modulated PWS (compatible with  \cite{CHAMBOULEYRON_NOISE_2023A&A...670A.153C}), $1$ for the modulated Sharp Bi-O-edge (close to the classical Zernike: 1.25) and $\sqrt{2}$ for the Grey Bi-O-edge which is remarkable provided that the Grey Bi-O-edge is a sensor conceived to have sufficient dynamic range  to be used in a stand-alone AO system.}

\correction{In Section, \ref{sec:MODAL_NOISE_C-MODEL},} the accuracy of the results obtained from the empirical model have been improved thanks to a model based on the work by \cite{Fauvarque_Kernel_2019JOSAA_36.1241F}. The C-model permits to show how the number of controlled modes come into play. Finally, results from an end-to-end model (Sect. \ref{sec:E2E_SIMULATIONS}) have confirmed the gain expected for the ELT SCAO configuration in the linear regime .

There are different directions to give to future works on an analytical and simulation point of view and regarding practical implementations.

The first important point is to develop the formalism and simulations for systems with large residuals in order to take into account Optical Gain \citep[e.g.][]{DEO_CLOSE_2021A&A...650A..41D, Chambou_Pyramid_2020A&A_644A_6C} and check if the advantages of the Bi-O-edge are conserved when the small phase regime is not met. 

\correction{We will also study the effect of WF discontinuities present in GSMTs,  \citep[e.g.][]{BERTOU_PETAL_2022A&A_658A_49B}.  Preliminary simulations indicate that the Bi-O-edge and the PWS behave similarly in presence of phase discontinuities. For example, both can measure petal errors which are smaller than the sensing wavelength but suffer from the phase wrapping problem for large amplitudes \citep{POURRE_LWE_2022A&A...665A.158P}. As well, we believe the impact of segments co-phasing residuals shall be similar to the PWS.}

On a fundamental point of view, the property expressed in Eq.\ref{H_mode_property} has even a more profound consequence than the one we identified on the sensitivity. The four times redundant H modes measurements in each quadrant can take advantage of the implementation of super-resolution \citep{Oberti_SUPER_R}. This will improve significantly the accuracy of very high orders by rejecting aliasing. Beyond the possibility to even control more modes classically allowed for a given resolution, this enrichment of the signal could be very beneficial to control non-linearity with advanced model-based WF reconstruction schemes \citep[e.g.][]{HUTTERER_2023InvPr..39c5007H} as well as with Machine-Learning techniques \citep[e.g.][]{Jalo_Pyramid_2022arXiv220507554N} and may help to solve some ELT related issues like the differential pistons issues \citep{BERTOU_PETAL_2022A&A_658A_49B}. Super-resolution will be the topic of a forthcoming paper.

Practically speaking, the Sharp Bi-O-edge can be seen as a mild evolution of the PWS concept with certainly an advantage for the manufacturing of accurate single-edge prisms. Its implementation would require only minimal developments. For XAO on an 8-m class telescope (say with 1000 correction modes) the Sharp Bi-O-edge would bring, with respect to the PWS, about $0.2 mag$ of gain and about $0.5 mag$ for XAO on an ELT. 

The advantage of the Grey Bi-O-edge is very significant, since the gain goes from $0.7 mag$ on an 8-m class telescope to $1.1 mag$ on the ELT for the current PCS baseline ($\approx 10^4$ modes) which may have a significant impact on the number of scientific targets available: in this case, to reach a similar AO performance than with a PWS, the Bi-O-edge can use $2.7$ times less photons. Hence it can use guide stars up to $\sqrt{2.7}$ times further away which corresponds to an observable volume that is more than $4$ times larger \correction{(we note that for $10^4$ corrected modes, the Sharp Bi-O-edge presents a gain of 1.6, corresponding to an observable volume about twice larger than the PWS)}. A forthcoming work will study in details the real impact on Science by evaluating S/N of coronagraphic images assisted by Bi-O-edge-based AO systems.

\correction{Apart from the high sensitivity, the absence of a tip-tilt modulation device presents an important advantage of the grey Bi-O-edge. In addition to the simplification of the design, the grey Bi-O-edge is not limited by the mechanical dynamics of a fast steering mirror and limited only by WFS camera and Real-Time Computer speeds. }

\correction{However, the complexity will be on the manufacturing of a (preferably) loss-less grey-scaled edge with a typical width of $100 \mu m$. One of the class of techniques we have thought about so far, is the beam splitting by  \textit{division of amplitude}. This can be done, for instance, by depositions of metal coatings of different depth, or by using dielectric plates, or by using rotators and polariser beam-splitters \citep[e.g.][]{GENDRON_YAW_2010aoel.confE5003G,Haffert_GODS_2016OExpr_2418986H,SNIK_V-APP_2012SPIE.8450E..0MS}. To deal with the variability of the mask and make use advantageously of micro-lithography techniques, we envision to define a discretisation of the amplitude. While waiting for detailed simulations,  we evaluate the need for a minimal value of $2$ resolution elements per $\lambda/D$ (for instance $12$ steps for a grey half-width of $3 \lambda/D$). This discretisation will also help with adjustments during the process and to deal with issues like amplitude-dependent phase shifts that are likely to occur. Among these solutions, one technique, patterned liquid crystal, has already been tested and even demonstrated on sky for the validation of a Generalised Optical Differential Sensor \citep{HAFFERT_GODSonsky_2018SPIE10703E..23H}. Even though the manufactured mask is significantly less demanding in terms amplitude variation than a grey FKE, this achievement is really remarkable and contributes to bringing polarisation techniques among our favourites.}

\correction{We believe that splitting the beam by \textit{division of wavefront} is certainly the cheapest and less risky solution. We can use a typical technique employed in coronagraphy by using micro-lithography with reflective microdots \citep{MARTINEZ_MIRCODOTS_2009A&A...495..363M}. Very high resolution (micrometers) can be obtained such that there is probably no need for a discretisation like the one mentioned for the division of amplitude. The division between reflected and transmitted beam can be made very clean and shall not introduce any amplitude dependent phase shift. However, the main drawback is that since the microdots must encode the desired focal plane amplitude (and not the intensity), the amount of reflected light and transmitted light is asymmetric by nature and leads to diffraction losses.}
\correction{Still, the microdots pattern could be optimised for the transmitted beam only (or reflected beam) which contains all the WF phase information. This concept would be simpler to implement opto-mechanically and may still be competitive in terms of sensitivity with the Sharp Bi-O-edge, but with the advantage of the static modulation.}

The overall opto-mechanical concepts for integrating two orthogonal FKEs need to be explored, especially for reaching sufficiently compact designs with minimal non common path aberrations.
\correction{The number of detectors is also an important topic. The Sharp Bi-O-edge has two channels, so two detectors is a logical solution. However, a smart design gathering all pupils on one detector should be possible without increasing Non Common Path Aberrations (NCPA). Indeed, since the WF is encoded into intensity signal at the level of the masks, only optics before the masks are contributing significantly to the  NCPA. The Grey Bi-O-edge has four channels, so four detectors is one potential solution. This may even be an advantage for very high order XAO and could allow a fine adjustment of pixels grid alignment for implementing super-resolution. Here also, smart designs may reduce the number of detectors needed. For instance, the polarisation technique could be based on transmissive optics, (Wollaston prisms, patterned Liquid crystal), allowing more compact designs.} 

To conclude, we believe that the Bi-Orthogonal-Foucault-knife-edge sensor with its outstanding capabilities in terms of sensitivity and resolution, is a timely new WFS candidate for the coming challenges in the field of HCI especially on GSMTs. 

\appendix
\section{Noise propagation with the C-model}
\label{sec:C_model}
\subsection*{Definitions}
\begin{itemize}[noitemsep]
    \item $x,y$: direct space variables
    \item $u,v$: Fourier space variables
    \item $d$:   Sampling size of CCD pixel projected on the telescope input pupil
    \item $f_c=\frac{1}{2d}$: sensor cut-off frequency
    \item $m_{\ma{wfs}}(u,v)$: focal plane amplitude filters
    \item $\omega(u,v)$: Modulation function
    \item $\ma{TF}_k(u,v)$: Quadrant $k$ Transfer Function
    \item $\phi(x,y)$: WF phase
    \item $M(x,y)$: Measurement function
    \item $M_a(x,y)$: Measure and average function
    \item $\Pi_d(x,y)=1, |x|<1/2, |y|<1/2$. Top-hat function
    \item $\circledast$: Convolution product
    \item $\hat{F}(u,v)$: Fourier Transform of $F(x,y)$
    \item ${.}^*$: complex conjugate
    \item $\ma{N}_{ph}$: incident flux in input pupil (photons/$m^2/frame$)
    \item $\Omega_{tel}$: Surface of the telescope ($m^2$)
\end{itemize}
The PWS is described by $4$ binary masks \citep[see][]{CARBILLET_CAOS_2005MNRAS.356.1263C}. We use the same definition and enumeration as in \cite{Fauvarque_Kernel_2019JOSAA_36.1241F} and for the Bi-O-edge, we extend it to the $4$ FKE masks represented in Fig. \ref{fig:mask_comparison_pyr_fke}. We express the masks in the Fourier space: 
\begin{itemize}
\item ${m_{pyr,1}:=1}$ for $u<0$ and $v>0$
\item ${m_{pyr,2}:=1}$ for $u>0$ and $v>0$
\item ${m_{pyr,3}:=1}$ for $u<0$ and $v<0$
\item ${m_{pyr,4}:=1}$ for $u>0$ and $v<0$
\end{itemize}
For the sharp Bi-O-edge, the masks definition is:
\begin{itemize}
\item $m_{bio,1}:=1$ for $u<0$
\item $m_{bio,2}:=1$ for $u>0$
\item $m_{bio,3}:=1$ for $v<0$
\item $m_{bio,4}:=1$ for $v>0$
\end{itemize}
By analogy with the PWS, we call 'quadrant' the pupil image resulting from the diffraction by a mask.

\subsection*{Sensitivity and Noise propagation}
In this section, we derive the equivalent of Eq. \ref{sigma_noise}. Thanks to the C-model \citep{Fauvarque_Kernel_2019JOSAA_36.1241F}, we can describe the sensitivity for any spatial frequencies. 

We define the noise propagation density in Fourier space:
\begin{equation}
\label{eq:NP_FOU}
\sigma_{\ma{wfs}}^2(u,v) =  \frac{\beta_{R,\ma{wfs}}^2(u,v)}{\ma{N}_{ph}}
\end{equation}
where $\beta_{R,\ma{wfs}}^2(u,v)$ defines the noise propagation function.
\subsection*{Transfer Function }
\label{sec:TF_MEAS}
The noise propagation function $\beta_{R,\ma{wfs}}$ is obtained from the TFs of the sensors. We call $\hat{M}_{\infty}(u,v)$ the meta-intensities in Fourier space with infinite spatial resolution. The TF is defined as:
\begin{equation}
\label{eq:M_TF}
\hat{\ma{M}}_{\infty}(\hat{\phi}(u,v)) = \ma{TF}(u,v) \hat{\phi}(u,v) 
\end{equation}
The TF for a \correction{generalized phase and amplitude mask} $m_{\ma{wfs},k}(u,v)$ is given by \citep{Fauvarque_Kernel_2019JOSAA_36.1241F}:
\begin{equation}
\ma{TF}_{\ma{wfs},k}=  i \left ( \hat{\hat{m}}_{\ma{wfs},k} \circledast (m_{\ma{wfs},k} \omega)^*  - m_{\ma{wfs},k}^* \circledast (\hat{\hat{m}}_{\ma{wfs},k} \omega) \right)
\label{eq:DEF_TF_MASK}
\end{equation}
where a circular modulation function $\omega(u,v)$ including diffraction is equal to:
\begin{equation}
\omega(u,v) = \delta (r-r_{mod}/D) \circledast \ma{PSF}_0(u,v)
\label{eq:DEF_O}
\end{equation}
where $r = \sqrt{u^2+v^2}$.
We use \correction{the pure amplitude masks representation} and the slope-like measurements definition (Eqs. \ref{eq:I_pyr_x}, \ref{eq:I_pyr_y} for the PWS and Eqs. \ref{eq:I_bio_x},\ref{eq:I_bio_y} for the Bi-O-edge)  and apply the C-model to get the TFs. The TF corresponding to the PWS slope-like definition is (correcting a sign in equation B13 of \cite{Fauvarque_Kernel_2019JOSAA_36.1241F}):
\begin{equation}
\label{eq:TFx_PYR}
\begin{aligned}
\ma{TF}_{x,pyr} = 2i [ m_{pyr,3} \circledast (m_{pyr,2} \omega) -  m_{pyr,2} \circledast (m_{pyr,3} \omega) \\ 
+ m_{pyr,1} \circledast (m_{pyr,4} \omega) -  m_{pyr,4} \circledast (m_{pyr,1} \omega) ]
\end{aligned}
\end{equation}

\begin{equation}
\label{eq:TFy_PYR}
\begin{aligned}
\ma{TF}_{y,pyr} = 2i [ m_{pyr,3} \circledast (m_{pyr,2} \omega) -  m_{pyr,2} \circledast (m_{pyr,3} \omega) \\ 
- m_{pyr,1} \circledast (m_{pyr,4} \omega) + m_{pyr,4} \circledast (m_{pyr,1} \omega)]
\end{aligned}
\end{equation}
A similar calculation gives the TF of the measurements for the Bi-O-edge:
\begin{equation}
\label{eq:TFx_BIO}
\ma{TF}_{x,bio} = 2i \left[ m_{bio,2} \circledast \left( m_{bio,1}{\omega}\right) 
- m_{bio,1} \circledast \left( m_{bio,2}{\omega}\right) \right]
\end{equation}
\begin{equation}
\label{eq:TFy_BIO}
\ma{TF}_{y,bio} = 2i \left[ m_{bio,4} \circledast \left( m_{bio,3}{\omega}\right) 
- m_{bio,3} \circledast \left( m_{bio,4}{\omega}\right) \right]
\end{equation}

\subsection*{Noise propagation in Fourier space}
\label{sec:WF_REC_FOU}
The sensitivity of a sensor depends on the resolution or pixel size $d$ (measured in the entrance pupil reference). The meta-Intensity $\ma{M}_{x}$ is the pixel-filtered version (same for $y$):
\begin{equation}
\label{eq:MEAS_inFOU_noSHIFT}   
\hat{\ma{M}}_x(\hat{\phi}(u,v)) = \hat{\ma{M}}_{a,x}(u,v) \hat{\phi}(u,v)
\end{equation}
where the Measure and average function $\hat{\ma{M}}_{a,x}(u,v)$ is given by (same for $y$):
\begin{equation}
\label{eq:MEASURE_AND_AVERAGE}   
\hat{\ma{M}}_{a,x}(u,v) = \ma{TF}_x(u,v)\ma{sinc}( d u) \ma{sinc}(d v)
\end{equation}
Following a similar development than \cite{Jolissaint_model2006JOSAA_23_382J}, we obtain the noise propagated:
\begin{equation}
\label{eq:NOISE_PROPAG_GEN}
\sigma_{\ma{wfs}}^2(u,v) = \frac{\sigma_{mI}^2}{\frac{1}{d^2}(|\hat{\ma{M}}_{a,x}(u,v)|^2 + |\hat{\ma{M}}_{a,y}(u,v)|^2  )} 
\end{equation}
We define the noise propagation coefficients as:
\begin{equation}
\label{eq:NOISE_PROPAG_COEF}
\ma{NC}_{\ma{wfs}}(u,v) = \frac{1}{\frac{1}{d^2}(|\hat{\ma{M}}_{a,x}(u,v)|^2+|\hat{\ma{M}}_{a,y}(u,v)|^2)}
\end{equation}
We define:
\begin{equation}
\label{eq:SIGMA_metaI_GEN}
\sigma_{mI}^2 =  \sigma_{N,\ma{wfs}}^2 = \frac{\gamma_{mI}}{d^2 \ma{N}_{ph}}
\end{equation}
$\gamma_{mI}$ is the result of the calculation of the theoretical variance of the meta-intensities $\sigma_{N,\ma{PYR}}$ and $\sigma_{N,\ma{PYR}}$. Finally we can write the function $\beta_R^2(u,v)$ as:
\begin{equation}
\label{eq:BETA_GEN}
\beta_R^2(u,v)=\frac{\gamma_{mI}}{|\hat{\ma{M}}_{a,x}(u,v)|^2+|\hat{\ma{M}}_{a,y}(u,v)|^2}
\end{equation}

\subsection*{Modal Noise Propagation}

We derive an approximate correspondence between spatial frequency and mode index.
With $f_r=\sqrt{u^2+v^2}$ the modulus of the spatial frequency we define an index called $i_{\ma{kl}}(f_r)$ that is the index of each mode (for instance KL modes). 
The lowest order modes corrected are tip and tilt and they correspond to the spatial frequency $f_{min} = \frac{1}{2D}$.
We write the effective order of the modes:
\begin{equation}
\label{eq:Order_F}
O_{f_r}= f_r / f_{min}
\end{equation}
We assume Noll's indexing convention holds \citep{Noll_Zernike_1976JOSA_66_207N} and write the KL index to frequency $f_r$ correspondence (including all azimuthal orders until $f_r$):
\begin{equation}
\label{eq:KL_Order}
i_{\ma{kl}}(f_r) = O_{f_r}(O{f_r}+1)/2+O_{f_r}
\end{equation}

The modal noise propagation coefficients from the C-model is obtained by averaging circularly Eq. \ref{eq:NOISE_PROPAG_COEF} and normalising by the telescope surface to get the noise variance per mode:
\begin{equation}
\label{eq:NC_Fr}
\ma{NC}_{\ma{wfs}}^{f_r} \approx \frac{<\ma{NC}_{\ma{wfs}}(u,v)>_{f_r}}{\Omega_{tel}}
\end{equation}

\bibliographystyle{aa.bst} 
\bibliography{BI-O-EDGE} 
\end{document}